\documentstyle[12pt,aaspp4]{article}

\begin {document}
\def\pedant{Mart{\'{\i}}n }
\title{ High resolution spectroscopy of ultracool M dwarfs}

\author {I. Neill Reid}
\affil {Space Telescope Science Institute, 3700 San Martin Drive, Baltimore, MD 21218 \\
Dept. of Physics \& Astronomy, University of Pennsylvania, 209 S. 33rd
Street, Philadelphia, PA 19104-6396; 
e-mail: inr@stsci.edu}

\author {J. Davy Kirkpatrick}
\affil {Infrared Processing and Analysis Center, 100-22, California Institute
of Technology, Pasadena, CA 91125}

\author {James Liebert}
\affil {Steward Observatory, University of Arizona, Tucson, AZ 85721}

\author {J. E. Gizis}
\affil {Department of Physics \& Astronomy, University of Delaware}

\author {C. C. Dahn, D. G. Monet}
\affil {U.S. Naval Observatory, P.O. Box 1149, Flagstaff, AZ 86002}

\begin {abstract}

We present high-resolution echelle spectroscopy of thirty-nine dwarfs with 
spectral types between M6.5 and L0.5. With one exception, those dwarfs were selected
from the 2MASS database using photometric criteria, (J-K$_S) \ge 1.1$ and 
K$_S \le 12.0$,  and therefore should provide a sample free of the kinematic biases which
can affect proper-motion selected samples. Two of the stars, 2MASSI 0253202+271333
and 2MASSW 0952219-192431, are double-lined spectroscopic binaries. 
We have used our observations to search for Li I 6708\AA\
absorption, characteristic of sub-stellar mass; estimate the level of chromospheric activity
through measurement of H$\alpha$ emission fluxes; measure rotational velocities via line broadening; 
and determine radial velocities and Galactic space motions. Two dwarfs have strong lithium absorption, 
the previously-known brown dwarf, LP 944-20, and 2MASSI J0335020+234235, which we identify as
a probable 0.06$M_\odot$ brown dwarf, age $\sim1$ Gyr. We have investigated the prospect of using
the observed frequency of lithium absorption amongst ultracool M dwarfs (M7 to M9.5) as a probe
of the initial mass function, comparing the observed frequency against predictions based on
recent theoretical models of low mass dwarfs and an assumed star formation history. Our results
show that the conclusions drawn are vulnerable both to systematic differences between the
available models, and to incomplete local sampling  of the most recent 
star formation events (ages $< 10^8$ years). The latter consideration stems from the mass-dependent rate of
evolution of brown dwarfs. Even given those caveats, however,
the available observations are difficult to reconcile with
Salpeter-like power-law mass functions ($\alpha \ge 2$) for masses below 0.1M$_\odot$. \\
A comparison between the rotational velocities and H$\alpha$ fluxes shows no evidence for 
significant correlation. The mean activity level of the ultracool dwarfs lies almost a factor of ten below that
of early- and mid-type M dwarfs. The relative number of dwarfs with $v \sin{i} < 20$ kms$^{-1}$
and $> 20$ kms$^{-1}$ is independent of spectral type. \\
Finally, velocity dispersions derived for our photometrically-selected sample of ultracool dwarfs 
are significantly lower than those measured for nearby M dwarfs, but 
show remarkable similarity to results for earlier-type emission-line (dMe) dwarfs. The
latter are generally assigned ages of less than $\sim3$ Gyrs.

\end{abstract}

\keywords{stars: low-mass, brown dwarfs; stars: luminosity function, mass function; 
 Galaxy: stellar content }

\section {Introduction}

The pioneering Two Micron Sky Survey (TMSS) undertaken by Neugebauer \& Leighton (1969)
provided the first large-scale celestial survey at near-infrared wavelengths. Extending
only to K$\sim3$rd magnitude and covering 75\% of the sky (Neugebauer, Martz \& Leighton, 1965), 
the TMSS revolutionised
our understanding of the nature of cool, luminous objects, such as Mira variables
(Ulrich {\sl et al.}, 1966), red supergiants (for example, VY CMa, Hyland {\sl et al.}, 1969) 
and dust-enshrouded AGB stars (for example, IRC 10216, Becklin {\sl et al.}, 1969). 
The current generation of surveys,
DENIS (Epchtein {\sl et al.}, 1994) and 2MASS (Skrutskie {\sl et al.}, 1997), reach flux levels more
than 10,000 times fainter than the TMSS, and provide the first census of the
near-infrared sky at moderately faint magnitudes. As such, they offer the prospect of
a similar revolution in our understanding of cool, low-luminosity
objects. Initial analyses have concentrated primarily on the coolest dwarfs discovered
in those surveys, sources whose atmospheric properties and consequent emergent energy 
distributions have necessitated the creation of new spectral classes: type L (Kirkpatrick {\sl et al.}, 
1999; \pedant {\sl et al.}, 1999) 
and, more recently, type T, for Gl229B-like dwarfs (Strauss {\sl et al.}, 1999; Burgasser {\sl et al.}, 1999).
The new surveys, however, also have the potential to revolutionise 
studies of late-type M dwarfs. 
Even if those objects are now a shade pass\'e, analyses of enlarged samples can
probe such fundamental parameters as the stellar mass function, $\Psi(M)$, 
kinematics and the origin and range of magnetic activity close to 
the hydrogen-burning limit. 

Photometric and spectroscopic data for nearby 
late-type M dwarfs are summarised in a series of papers by Kirkpatrick {\sl et al.} (1995, 1997)
and Henry {\sl et al.} (1995, 1997). That sample includes 26 dwarfs with spectral types between
M7 and M9.5, which, following Kirkpatrick {\sl et al.} (1995), we
designate ultracool M dwarfs. A minority of those sources were identified in 
deep photometric surveys by Reid \& Gilmore (1981), Kirkpatrick {\sl  et al.} (1994),
Tinney {\sl et al.} (1993) and Kirkpatrick {\sl et al.} (1997). The majority, however,
were discovered through spectroscopic follow-up observations of faint red stars
from proper motion surveys, primarily Luyten's Palomar catalogues. As a result,
a statistical analysis runs the risk of bias due to preferential
inclusion of higher-velocity stars, which are more likely to be drawn from
the older stars in the Galactic disk, while young, low space-motion objects, 
such as M-type brown dwarfs, lie undetected. The availability of near-infrared
survey data provides the first opportunity for the construction of a substantial
sample of ultracool M dwarfs using purely photometric criteria.

This paper presents high-resolution spectroscopic observations of 38 bright, ultracool M
dwarfs and one L dwarf drawn from the 2MASS database. Our analysis has three main goals: first, an
estimate of the fraction of ultracool dwarfs which exhibit detectable lithium absorption;
second, the derivation of improved statistics on the distribution of rotational velocities; and,
finally, a determination of the kinematics of these low luminosity dwarfs.
We have also used the echelle spectra to study the range of chromospheric activity 
at these spectral types.
The paper is structured as follows: section 2 describes sample selection and the 
spectroscopic observations; section 3 summarises the results of our search for lithium 
absorption; section 4 discusses our
measurements of rotational velocities and comments on correlations  with the level
of chromospheric activity; section 5 considers the space velocity distribution; 
and section 6 presents our conclusions.

\section {Sample selection and spectroscopic observations}

\subsection {Sample selection}

Achieving the scientific goals outlined in the introduction requires high
signal-to-noise observations at high spectral resolution of late-type M dwarfs. To that
end, we have selected our targets from the 2MASS database using both colour and
apparent magnitude criteria. 2MASS provides photometry in three passbands:
J(1.25$\mu m$), H(1.6$\mu m$) and K$_S$ (2.2$\mu m$), where the subscript `S' 
denotes that the passband is truncated at the long wavelength (thermal) limit
as compared with the standard K passband (see Persson et al, 1999). 

Our selection criteria are derived from the (M$_J$, (J-K$_S$)) colour-magnitude diagram
described by nearby stars and brown dwarfs with accurate (${\sigma_\pi \over \pi} < 15\%$)
trigonometric parallax  measurements (Figure 1). The photometric data plotted in this figure 
are from Leggett (1992), the 2MASS database and Dahn {\sl et al.} (2000). M dwarfs
outline an almost-vertical main sequence with 0.8 $< \langle$(J-K$_S) \rangle < 0.9$
to M$_J \sim 10.5$, corresponding to spectral types earlier than $\approx$M6.
At lower luminosities, the main sequence moves redward: GJ 1111 (spectral type M6.5) has
(J-K$_S$)=1.04; VB8 (M7) has (J-K$_S$)=1.05; VB10 (M8) has (J-K$_S$)=1.10; and
LHS 2924 (M9) has (J-K$_S$)=1.17 magnitudes. 
(J-K$_S$) continues to increase to a maximum value of $\sim2.1\pm0.1$
magnitudes for the latest-type L dwarfs, such as Gl 584C (L8, Kirkpatrick {\sl et al.}, 2000).

Our observations are aimed at ultracool M dwarfs, with spectral types between M7 and M9.5.
Following Figure 1, the majority of our targets are chosen to have near-infrared colours in the range
$1.3 \ge {\rm (J-K}_S) \ge 1.1$ and K$_S \le 12.0$ magnitudes. Table 1 lists positions, IJHK$_S$
photometry and spectral types for our sample. The 2MASS K$_S$ magnitudes have typical accuracies
of $\pm$0.01 - 0.02 magnitudes, and the near-infrared colours have typical uncertainties of $\pm0.03$
magnitudes. Most of the Cousins I-band magnitudes have been synthesised from flux-calibrated low-resolution
spectra, obtained at either the Keck Observatory or at Las Campanas Observatory (see Gizis {\sl et al.}, 
2000a, for details). Comparing these measurements against broadband data for stars with conventional
photometry indicates uncertainties of $\pm0.1$ magnitude. 

Four dwarfs listed in Table 1 require comment: 
2MASP J1242464+292619, or 2M1242+29 (we use this abbreviated form hereafter), 2M0339-35, 
2M0149+29 and 2M2234+23. The last two are M9.5 dwarfs which have (J-K)$_S > 1.3$; both were
identified as candidate L dwarfs. 2M1242+29 was identified in 2MASS Protocam JHK data 
(Kirkpatrick {\sl  et al.},
1997), and is significantly fainter than our formal magnitude limit, with
K$_S=13.23$. While we tabulate our observations of this star, we have not included it in any of
the statistical analyses discussed further below. Finally, 2M0339-35 (LP944-20) meets our selection
criteria, and was observed with HIRES, but inclement conditions (due to the high airmass)
led to a low signal-to-noise spectrum. Fortunately, 
previously-published observations are available for this dwarf (Tinney, 1998; 
Basri, 2001), and we use those data in our analysis.  

Even excluding 2M1242+29, the dwarfs listed in Table 1 do not constitute a 
complete, magnitude-limited sample. However, they are a randomly-chosen subset of
2MASS sources with the specified magnitudes and colours, and therefore should 
provide a representative sample (specifically, kinematically unbiased) of nearby 
ultracool dwarfs. Several of the stars in Table 1 are well-known late-type dwarfs
with moderate or high proper motions; the remainder have been confirmed as ultracool dwarfs
through intermediate-resolution spectroscopy, mainly by 
Gizis {\sl  et al.} (2000a; hereinafter, G2000). The thirty-nine dwarfs include fourteen with spectral types
M7 or M7.5, fourteen of type M8/M8.5 and seven at M9/M9.5. Three of the remaining four dwarfs have
spectral types M6/M6.5; the last is the brightest L dwarf currently known, 2M0746+20(L0.5), 
recently resolved as a near equal-mass binary through HST observations (Reid {\sl et al.}, 2001).

Figure 2 plots the JHK two-colour diagram for the sample, where we 
identify the different spectral types. The (J-K$_S$) selection limit is apparent
from the distribution.  We use data for the parallax sample plotted in Figure 1 
to outline the characteristic tilted-S shape of the main-sequence.
The turnover in (J-H) at (H-K$_S)\sim0.2$
stems from a combination of two factors: the wavelength dependence of H$^-$ opacity, which
peaks at $\sim1.6 \mu$m, and a shallower temperature gradient as convection becomes more important
at spectral types later than M1/M2 (Mould, 1976). The higher opacity 
places the photosphere at shallower depths and lower temperatures, reducing the total flux 
emitted in the H passband. The blueward trend reverses at spectral
type $\approx$M5, as H$^-$ becomes a less important opacity source at lower temperatures, 
and (J-H) increases rapidly at later spectral types. Kirkpatrick {\sl  et al.}
(2000) discuss the near-infrared colours of ultracool M and L dwarfs, and comment that
the near-infrared absolute magnitudes are better correlated with spectral type 
than with (J-K$_S$). The corresponding dispersion is evident in Figure 2: the three M6/M6.5
dwarfs are significantly redder than most of the M7/M7.5 dwarfs, while 2M0746+20 is not
the reddest dwarf in the sample. 

\subsection {HIRES spectroscopy}

High-resolution spectroscopic data were obtained of all targets using the 
HIRES echelle spectrograph on the Keck I telescope (Vogt et al, 1994). The
observations were made on November 14 (UT), 1998, March 8, 
June 14-16 and December 29-30, 1999. Conditions were good throughout these runs, 
with clear skies and seeing between 0.7 and 0.9 arcseconds. We used a
slit width of 0.86 arcsecond for all observations, giving data of resolution 
2.9 pixels or R$\sim 33000$ (8.9 kms$^{-1}$). While there are slight differences
in the grating angles from run to run, our spectra span
the wavelength range 6300 to 8600\AA. Coverage is
not continuous, due to the physical size of the $2048^2$ Tektronix CCD, but
all of the spectra include the H$\alpha$ line, Li I 6708\AA, the
7050\AA\ TiO bandhead, the K I 7665/7699\AA\ doublet and the 8432\AA\ TiO bandheads.

Basic data reduction was undertaken using the suite of programs written by
T. Barlow. The spectra were corrected for the instrumental response using
flat fields taken with an internal tungsten lamp, and an optimal extraction
algorithm applied to extract the individual orders in each frame. The wavelength
scale is defined by thorium-argon arc-lamp exposures, taken at the beginning, 
middle and end of each night. 
We used  standard {\sl iraf} routines to determine the calibration, and intercomparison
shows that the zeropoint is stable to $\pm1$ kms$^{-1}$ on any given night. 

\subsection {Chromospheric emission}

All of the dwarfs observed possess emission lines indicative of
chromospheric activity, although for LHS 2632 the emission lies at the threshold of
detection, even with HIRES.
The most prominent features are the H$\alpha$ lines, although many dwarfs
also exhibit emission in the cores of the KI 7665/7699 doublet, while approximately
25\% of the sample show noticeable He I 6678\AA\ emission. Weak He I emission
(EW $<$ 0.3\AA) may be masked in other dwarfs by the complex TiO absorption in
that part of the spectrum. 

Table 2 lists the equivalent widths measured for the hydrogen and helium lines,
and Figures 3 to 6 plot our spectra in the vicinity of the H$\alpha$ line.  
We also list the activity level, $F_\alpha \over F_{bol}$, discussed further in
\S 4.
Two dwarfs stand out: 2M0320+18 (LP 412-31), with an H$\alpha$ equivalent width of 83\AA, 
and 2M0350+18, where H$\alpha$ reaches an equivalent width of almost 40\AA. As with the
rest of the sample, these two dwarfs have previous spectroscopic measurements at
these wavelengths, albeit at lower resolution. Gizis {\sl  et al.} (2000a) measure an 
H$\alpha$ equivalent width of 29\AA\ in the former case, but failed to detect any
emission from 2M0350+18. Evidently, both of the current observations were made while these
two stars were undergoing strong flares.

Conversely, three dwarfs have significantly lower levels of activity in our spectra
than measured in previous observations. Observations of 2M0149+29 during a substantial outburst
are described by Liebert {\sl et al.} (1999), while both LHS 2243 and LP 475-855 were
significantly more active at the time of the low resolution observations described in G2000. 
These variations and the overall distribution of activity are 
discussed in more detail in \S 5.

\subsection {Radial velocities}

We have employed two methods to determine radial velocities for the dwarfs in the
present sample. First, we measure the central wavelength of H$\alpha$ emission
and correct the apparent radial velocity for heliocentric motion using the {\sl iraf}
routine {\sl vhelio}. Those measurements are listed in column 5 of Table 2 as V$_\alpha$.
In most cases, we estimate measurement uncertainties of $\pm$2 to 3 kms$^{-1}$; the exceptions
are dwarfs with weak emission or broad H$\alpha$ profiles, such as LHS 2632, 2M1047+40A and 
TVLM 513-46546. 

We have also determined radial velocities using standard cross-correlation routines. Given
the seasonal variation in observing runs, and slight variations in the
instrumental setup, we were unable to use the same  radial
velocity standard for all of the targets. Our observations are tied
to three reference stars: LHS 2065 (M8) in March, 1999; Gl 412B (M5) in June, 1999; and
Gl 83.1 (M4.5) in December. Marcy \& Benitz (1989) measure a heliocentric radial velocity
of -28.6 kms$^{-1}$ for the last star, while Delfosse {\sl  et al.} (1998) cite
V$_{hel} = 68$ kms$^{-1}$ for both components of Gl 412. Finally, Tinney \& Reid (1999)
measure V$_{hel} = 8.7\pm1.5$ kms$^{-1}$ for LHS 2065.

Both Gl 83.1 and Gl 412B have earlier spectral types than the dwarfs in our sample.
However, the cross-correlation peaks generally exceed 0.5 in height in orders dominated by
TiO absorption, rather than by strong atomic absorption or emission lines.
The velocities listed in column 6 of Table 2 (V$_{CCF}$) are averaged from measurements
of six to nine echelle orders, with the uncertainty reflecting the rms dispersion about the mean.

In general, there is good agreement between the radial velocities derived using the
two methods. However, several dwarfs exhibit anomalies. Two are clearly binary: 
as Figure 7 shows, 
the cross-correlation spectra for both 2M0253+27 and 2M0952-19 exhibit two 
distinct peaks, characteristic
of double-lined spectroscopic binaries. In the former case, the cross-correlation peaks have
heights in the ratio 4:1 at 8000\AA, with the blue-shifted component the stronger of the two; in
the latter, the flux ratio is 2:1, again in favour of the component with the more negative velocity.
We list CCF velocities for both components
of those dwarfs in Table 2. Neither shows evidence for separable H$\alpha$ emission
(Figures 3 \& 4), so 
V$_\alpha$ presumably reflects a weighted average of the centroids of the individual
emission lines. We note that 2M1550+30 shows an asymmetric cross-correlation peak
(Figure 7), suggestive of an SB2 binary with a secondary close to the detection limit 
(as in figures 5 and 6, Reid \& Mahoney, 2000; hereinafter, RM2000). 

Four other dwarfs require comment: 2M0350+18, 2M1047+40B, 2M1336+47 and 2M2206-20. In each case, V$_\alpha$
and V$_{CCF}$ differ by more than 5 kms$^{-1}$. The last three dwarfs are all fast rotators,
with broad cross-correlation peaks and asymmetric H$\alpha$ profiles (Figure 3 to 6). The
observed discrepancies probably arise in our estimation of the centroid of the H$\alpha$
line, and hence of V$_\alpha$. 2M0350+18, in contrast, is a slow rotator, but has an unusual
H$\alpha$ profile, with a narrow spike superimposed on a broad pedestal. This is
reminiscent of one observation of the Hyades SB2 binary, RHy 42 (RM2000). In that case, the
unusual morphology arises from the superposition of two normal H$\alpha$ profiles at a velocity
separation of $\Delta V \sim 50$kms$^{-1}$. Our present data, however, show no evidence for 
binarity on the part of 2M0350+18. The H$\alpha$ profile of this star is discussed further below.
In subsequent sections we adopt V$_{CCF}$ as the reference radial velocity for all the dwarfs 
in our sample. 

As noted above, the repeatability of the thorium-argon calibration spectra 
indicate that our velocity calibration should be accurate to $\pm1$ kms$^{-1}$; 
this is generally supported by analysis of the night-sky emission spectra extracted
from our data (as discussed in more detail by RM2000). 
We have a limited number of external tests of our measurements. 
Two late-type M dwarfs in the present sample were also observed by Tinney \& Reid
(1999): they measured radial velocities of  
V$_{hel} = -0.6\pm2$ kms$^{-1}$ for BRI1222 and V$_{hel} = 8.1\pm3$ kms$^{-1}$ for TVLM 513-46546;
we measure velocities of $-5.6\pm0.4$ kms$^{-1}$ and $5\pm6$ kms$^{-1}$, respectively. 
TVLM 513-46546 is one of the most rapidly rotating dwarfs in the present sample, as discussed
further below, accounting for the large uncertainty in V$_{CCF}$. The 5 kms$^{-1}$ offset
for BRI1222 exceeds the combined formal uncertainties of both measurements. Further
observations are required to settle the discrepancy; we apply no adjustment to our
measured velocities. 

\subsection {Rotation}

Tonry \& Davis (1979) demonstrated that the width of the peak of a cross-correlation
function is dependent on the line profiles present in the template and program object. 
In the case of stars, the dominant contributor to line-width is usually rotational
broadening, allowing measurement of the projected rotational velocity. The measured
full-width half-maximum of the CCF peak
is calibrated against  $v \sin{i}$  by applying artificial broadening for a
range of velocities to the spectrum of a slowly rotating star.
We have adopted the line-profile prescription given by Gray (1982) and  use
this technique to estimate $v \sin{i}$ for the dwarfs in the present sample. 

Our analysis generally follows the procedures described by RM2000.
We have limited analysis to spectral orders spanning the wavelength range
$\lambda \lambda 7360 -  7460$\AA\ and $\lambda \lambda 7840 - 7960$\AA, both covering
regions which lack TiO bandheads, strong atomic lines and significant terrestrial
absorption. The resulting measurements, using Gl 83.1, Gl 412B  and LHS 2632 as templates,
are given in Table 2.  The intrinsic resolution of the HIRES data corresponds to a rotational
velocity of $v \sin(i) \sim 2.5$ kms$^{-1}$; however, 
both Gl 83.1 and Gl 412B are known to have higher rotational velocities (Table 2), and, as a result, 
our detection limits are effectively  $v \sin(i) \sim 4$ kms$^{-1}$ (12/99 data) and
$v \sin(i) \sim 6$ kms$^{-1}$ (6/99), respectively. Those limits are  still sufficient to
distinguish rapid rotators from more conventional late-type M dwarfs. Our third
radial velocity standard, LHS 2065, has a modest
rotational velocity (Basri, 2001), so we have used the slow rotator, LHS 2632, as the rotational
reference for the March 1999 observations.  We estimate velocity
uncertainties of $\pm2$ kms$^{-1}$ at low $v \sin{i}$ and $\sim5$ kms$^{-1}$ for the fast
rotators. 

Basri (2001) has recently published rotational velocity measurements for more than 70 dwarfs with
spectral types later than M5. The majority of these measurements are 
based on Keck HIRES data, and therefore have similar resolution to our dataset.
There are nine dwarfs in common with our sample, and Table
3 compares the two sets of results. In general, the agreement is within the expected
uncertainties. The exception is
2M1224-12 ($\equiv$ BRI1222), where our measurement indicates modest rotation of 8 kms$^{-1}$, as compared with
only 2.0 kms$^{-1}$ detected by Basri. Table 3 also shows that the star was significantly more active
at the time of our observation. These results are discussed in more detail in \S4. 

\section { Lithium in ultracool M dwarfs}

 A major goal of this  project is determining the fraction of late-type M dwarfs in the Solar Neighbourhood
which exhibit lithium absorption. As originally discussed by Rebolo {\sl et al.} (1992),
lithium fusion (Li$^7$ + $p \rightarrow$ He$^4$ + He$^4$ ) requires a 
temperature of $T_{Li} \sim 2 \times
10^6$K. Since hydrogen fusion occurs at $T > 3.5 \times 10^6$K, primordial lithium
is destroyed rapidly in fully convective low-mass stars and
higher-mass brown dwarfs. Objects with masses below a specific mass limit, $M_{Li}$, 
are predicted to have central temperatures lower than the critical threshold, T$_{Li}$, 
even when evolving through the M dwarf r\'egime, and these objects preserve lithium at its
primordial abundance. The exact mass limit is model dependent, with current estimates of $M_{Li}$ ranging from
0.06 to 0.065 M$_\odot$ (Chabrier \& Baraffe, 1997; Ushomirsky {\sl et al.}, 1998); 
in any case, detection of Li 6708\AA\
absorption in an L dwarf is a clear indication of substellar mass. 

At higher masses, $M > 0.065 M_\odot$, the rate of lithium depletion increases with
increasing mass. This correlation permits the use of lithium detection as a mass discriminant in
open clusters and the general field. Figure 8 plots theoretical tracks for low-mass
stars and brown dwarfs, drawn from both the Burrows {\sl  et al.} (1997: the Arizona models) and the
Baraffe {\sl  et al.} (1998: the Lyon models) sequences. The two sets of model calculations show
good agreement for $M < 0.06 M_\odot$, but differ at higher masses. In particular, the Lyon models
place the hydrogen-burning limit at $\sim0.072M_\odot$, while a 0.075 $M_\odot$ dwarf is a transition
object in the Arizona calculations. The implications of these differences for the present analysis
are discussed further below. For both sets of models, Figure 8 outlines the lithium-depletion locus,
defined where the lithium abundance is predicted to
drop to 1\% of the primordial value.
The critical effective temperature ranges from 3200K for a 0.1$M_\odot$ star to 
$\sim2600$K for a 0.065 $M_\odot$ brown dwarf. 

Spectral type is correlated primarily with effective temperature. Thus, Figure 8 can be used to predict
the spectral type where lithium should become detectable in young and
intermediate-age star clusters; alternatively, the observed spectral type 
(or luminosity) at the
threshold for lithium detection can be used to infer the age of the cluster.
Stauffer, Schultz \& Kirkpatrick (1998) use the latter approach to
estimate an age of 120 Myrs for the Pleiades, while Stauffer {\sl et al.} (1999)
derive an age of 90 Myrs for the $\alpha$ Persei cluster.
M dwarfs in the general field span a wide range of ages. However, both sets of models 
plotted in Figure 8 indicate that the lithium depletion line meets the $M = M_{Li}$
 evolutionary track at a temperature of $\sim2600$K and an
age of $\tau \sim 400$ Myrs; that is, all dwarfs which deplete lithium 
have completed the depletion cycle by this age/temperature. As a corollary, any
dwarf with an inferred effective temperature lower than 2600K and detectable lithium
absorption is predicted to be a brown dwarf with a mass less than 0.065$M_\odot$.

This characteristic behaviour allows us to probe the stellar mass function. The
relative number of ultracool dwarfs with and without lithium absorption in our
sample, $F_{Li}$, depends on the relative number of dwarfs with masses $M < 0.065 M_\odot$ 
and $M > 0.065 M_\odot$.  Thus, if the immediate Solar Neighbourhood provides a
fair sampling of the disk population, if our sample is an unbiased subset of
local dwarfs, and if the stellar birthrate, B(t), is well behaved, then $F_{Li}$ depends on the 
shape of the mass function, $\Psi(M)$: the steeper $\Psi(M)$, the more low-mass dwarfs and
the higher $F_{Li}$. As discussed further below, circumstances may prevent us from satisfying all
the requisite conditional statements.

\subsection {Lithium detections in the present sample}

Our observations were designed specifically to include coverage of the Li 6708\AA\
doublet, the strongest feature due to that species. Figures 9 to 12 plot our
data for that region of the spectrum, marking with the expected location of the Li I
doublet, adjusted for the apparent velocity at the time of observation.  
Table 2 lists the results: of the
39 ultracool dwarfs in the sample, two show unequivocal absorption at that
wavelength, with one other possible detection. 
2M0339-35 (LP 944-20) was
known as an M-type brown dwarf prior to our observations (Tinney, 1998). As Figure 9
shows, 2M0335+23 stands out from the other dwarfs in the present sample, with a
broad, distinct feature at the appropriate wavelength. This dwarf
is strikingly similar to
2M0339-35, with comparable spectral type (M8.5 $vs.$ M9), nearly identical rotational
velocity ( $v \sin{i} = 30$kms$^{-1}$ $vs.$ 28 kms$^{-1}$) and lithium
absorption of comparable strength.
Our data suggest that 2M0335+23 is chromospherically more active, although
the difference is within the range of variation for individual dwarfs (see \S4). Given
those characteristics, we identify 2M0335+23 as a brown dwarf, with a likely age of $\sim1$ Gyrs and
a mass of 0.06 $M_\odot$. 

The possible lithium detection is for the M8 dwarf, 2M1242+29. As noted
above, this is fainter than the magnitude limit of our photometric sample, and
the HIRES spectrum is correspondingly noisy. Nonetheless, there appears to be
a relatively broad absorption feature at the appropriate wavelength for Li I 6708. 
This dwarf was discovered from observations undertaken with the prototype 
2MASS camera (Kirkpatrick, Beichman \& Skrutskie, 1997), and the Palomar spectrum
obtained at that time does not have sufficient resolution or signal-to-noise to
detect this relatively weak feature. If confirmed by further observations, the
measured equivalent width suggests little lithium depletion, and indicates a
mass below 0.065$M_\odot$ for 2M1242+29.

Considering the sample as a whole, our observations indicate that 36 of the 39 ultracool 
dwarfs have masses above the lithium depletion limit.
2M0335+23, 2M0339-35 and possibly 2M1242+29 are 
identified as brown dwarfs with masses below  $0.065_\odot$.
Of these three dwarfs, only 2M0335+23 and 2M0339-35 are included in the
photometrically-selected sample. 

\subsection {Modelling the lithium fraction}

Deriving $F_{Li}$ is a straightforward observational process. Subdividing our sample by
spectral type, $F_{Li} = 6\pm4\%$ for spectral types M7 to M9.5 (2 of 35 dwarfs), and
 $F_{Li} = 10\pm7\%$ for spectral types M8 to M9.5 (2/20). Interpreting
those measurements in terms of the shape of the mass function, $\Psi(M)$, is
more complicated, and requires use of the models plotted in Figure 8. 

First, we require a relation between spectral type and effective temperature. 
As discussed elsewhere
(Reid {\sl  et al.}, 1999a - R99; Kirkpatrick {\sl  et al.}, 2000), this
calibration remains uncertain, largely due to the complex nature of atmospheres at
these cool temperatures. Leggett {\sl  et al.'s} (1996) multiwavelength analysis
sets a benchmark of T$_{eff} = 2700$K at spectral type M6.5 (GJ 1111), while current concensus
places the boundary between the M and L spectral types at T$_{eff} = 2050\pm50$K 
(Kirkpatrick {\sl  et al.}, 2000; Schweizer {\sl et al.}, 2001, 2002). Given
those results, we adopt the following temperature/spectral type relations:
\begin{description}
\item[M7 - M9.5:] $2700 \ge T_{eff} > 2050$
\item[M8 - M9.5:] $2500 \ge T_{eff} > 2050$
\end{description}

Our goal is to compare predicted and observed values of $F_{Li}$ for a 
range of assumed mass function. We follow the techniques outlined by R99:
Monte Carlo simulations are used to generate a sample of `dwarfs' with
known distance, $d$, and age, $\tau$; masses are drawn from a power-law mass function,
$\Psi(M) \propto M^{-\alpha}$; we assume a uniform birthrate, B(t), 
with $10^{10} > {\rm t} > 10^7$ years, where t is lookback time (i.e. t=0 at the
present epoch).
Given $M$, $\tau$ and $d$, we compute m$_{bol}$ and T$_{eff}$ from the theoretical tracks, and, 
using the appropriate bolometric corrections (R99), calculate J, H and K$_S$ magnitudes. 
Based on
those data, we identify dwarfs with $K_S \le 12.0$ and temperatures in the relevant range,
and compute the fraction predicted to have undepleted lithium.

The most significant complication in interpreting these model predictions is
illustrated by Figures 13 and 14, which show simulated (mass, T$_{eff}$) and (mass, age) 
distributions
of ultracool dwarfs with K$_S < 12$ drawn from a power-law mass function with $\alpha=1$.
With no internal energy source, brown dwarfs
`cool like a rock' (Burrows \& Liebert, 1993), with the rate of cooling increasing with
decreasing mass. Selecting a sample based on a fixed range in spectral type
(temperature) is therefore equivalent to selecting dwarfs within a particular range of ages
for a given mass. In particular,
low-mass brown dwarfs (LMBDs - $M < 0.03 M_\odot$) cool so rapidly
that most enter the L dwarf r\'egime by $\tau \sim 10^8$ years (Figure 8); that is, 
LMBDs are only eligible for inclusion in an ultracool M dwarf sample for $< 10^8$ years. 
As a result, our conclusions are vulnerable to systematic bias introduced either by
the assumed stellar birthrate, or by age-specific inhomogeneities in the Solar 
Neighbourhood disk population.

Considering the stellar birthrate, we adopt the simplest assumption of a uniform
star formation over the history of the disk. We considered this issue in our
analysis of the sub-stellar mass function (R99), since 
the observed numbers of field L and T dwarfs are also dependent on the
recent star formation history. We concluded that the available data, based
primarily on the distribution of Ca II activity amongst G dwarfs
(for example, Soderblom, Duncan \& Johnson, 1991), were
consistent with a uniform birthrate. More recently, Gizis {\sl et al.} (2002, PMSU3) have 
re-examined this issue, and find that the distribution of chromospheric activity amongst 
nearby M dwarfs is also consistent with a relatively uniform age distribution.
These results are in contrast to studies of the global star formation history, 
which tend to favour star formation rates that
increase by an order of magnitude between the present epoch and redshift $\sim 1.5$
(Madau,  Pozzetti \& Dickinson, 1998). We note that if such a history were appropriate 
to the Galactic disk, then we would expect a higher proportion of long-lived, {\sl bona-fide}
stars amongst our ultracool sample. 

Spatial inhomogeneities are a potential problem since young stars are
not a well-mixed population, but tend to lie
in or near star-forming regions. Such regions are absent from the immediate Solar 
Neighbourhood, 
suggesting that a local sample, such as that discussed here, might be deficient in such
youthful objects relative to a global average over the disk. That deficiency manifests
itself as fewer lithium-strong brown dwarfs, which, in turn, could lead to our underestimating 
$\alpha$, the power-law index of $\Psi(M)$. On the other hand, there are some extremely young 
stars in the Solar Neighbourhood, notably the TW Hydrae association, $\tau \sim 2 \times 10^7$ years
(Kastner {\sl et al.}, 1997).  Moreover, approximately 1\% of the 
G dwarfs in Henry {\sl et al.'s} (1996) Ca II survey
have  activity levels consistent with ages of less 10$^8$ years (Soderblom, King \& Henry, 1998). 
With distances of less than 50 pc., these stars are drawn from the same volume as
our ultracool dwarf sample, although one should also note that, with a total sample
of $\sim800$ stars, subdividing on such a fine timescale leads to  correspondingly high 
statistical uncertainties.
In any case, we allow for possible age bias in the ultracool sample 
by applying several lower age limits, $\tau_{min}$, in the model calculations. 

\subsection {Results: constraints on $\Psi(M)$}

Table 4 summarises the results of our simulations. As noted above, 
we consider two observational samples:
spectral types M7 to M9.5, $F_{Li} = 6\pm4\%$; and spectral types M8 to M9.5,  
$F_{Li} = 10\pm7\%$. The table lists predictions based on
both the Arizona and Lyon theoretical tracks, and for mass-function indices
$0 \le \alpha \le 2$. We adopt the age/mass lithium depletion limits outlined
in Figure 8 - 0.065 M$_\odot$ for the Arizona models, and 0.07 to 0.06 M$_\odot$ for 
the Lyon dataset. Relatively few brown dwarfs lie in this mass range, 
so changing those limits by $\pm0.005 M_\odot$ has little effect on the results. 
The predicted percentage of dwarfs with detectable lithium are listed for 
lower age limits of 10$^7$, $5 \times 10^7$ and 10$^8$ years. 

The models allow us to quantify some issues raised in the introduction to this
section. First, 
dwarfs with $M > M_{Li}$ make little contribution to $F_{Li}$ in either
set of calculations. Even for the flattest mass function, $\alpha = 0$, higher-mass,
partially lithium-depleted brown dwarfs comprise less than 0.5\% of the total, and
the contribution of those objects becomes negligible for $\alpha \ge 1$.  Thus, absent
other considerations,
the lithium fraction provides a clean estimate of the slope of the underlying mass function.

Second, the quantitative results confirm the strong dependence of $F_{Li}$ on the effective
lower age-limit, $\tau_{min}$,  of the local sample. The systematic bias increases as the underlying
mass function steepens, and the relative number of young LMBDs increases. 

It is also clear that the two sets of models make significantly different
predictions of $F_{Li}$ for a given value of $\alpha$: 
the Lyon models predict lithium fractions which are lower by almost a factor of two
than those predicted by the Arizona models. As noted above, this reflects differences
in modelling the stellar, rather than brown dwarf, r\'egime. Figure 8 shows that
the Lyon 0.075 and 0.09 $M_\odot$ Lyon models essentially bracket the M7-M9 temperature
range for $\tau > 2 \times 10^9$ years. In contrast, that
region of the (T$_{eff}$, $\tau$) plane is populated by a more restricted range of
masses, $\sim 0.078$ to $\sim 0.088 M_\odot$, in the Arizona models. As the histograms
in Figures 13 and 14 illustrate, the result is that the
latter simulations predict fewer stellar-mass M7 to M9 dwarfs, and 
a correspondingly higher fraction of lithium-rich ultracool M dwarfs.

The correlation with $\tau_{min}$, and the systematic disagreement
between the two sets of models complicate the interpretation of the observed value of
 $F_{Li}$ as a constraint on $\Psi(M)$. Moreover, 
quantifying the comparison between the models and observation is difficult, not least
because all of the relevant observational parameters (our estimates of $F_{Li}$ and
the young G-dwarf fraction) have significant associated uncertainties. As a
first cut, we assume a moderate bias against very young objects locally ($\tau_{min} = 0.05$ Gyr.), 
and consider
the constraints set if we require the predicted value of the lithium fraction fall
within 1$\sigma$ of the observations. 

Under those criteria, the Lyon models suggest that a power-law mass function
is consistent only for $\alpha < 1.5$, while 
predictions based on the Arizona models require a significantly flatter mass function, 
$\alpha \le 0.5 $. The different indices reflect the relative contribution of stars
and brown dwarfs to the ultracool sample; as discussed above, the 
Arizona models predict larger numbers
of sub-stellar mass objects, and, as a result, require a flatter index to match the
low lithium fraction  in the observed sample. The derived indices are generally
consistent with previous analyses, both based on the surface densities of L and
T dwarfs in the field ($\alpha \sim 1.3$, R99), and based on
surveys of young clusters (Luhman {\sl et al.}, 2000; Luhman, 2000; 
Barrado y Navascu{\' e}s {\sl et al.},  2001). In general, the open cluster analyses,
where age is less of an uncertainty,
favour flatter mass functions, with $0.5 < \alpha < 1$. 

What do these estimates imply for the local density of brown dwarfs? Modelling the
stellar mass function as a power-law, $\alpha$=1, for $1.0 > {M \over M_\odot} > 0.075$,
then brown dwarfs are predicted to outnumber stars by $\sim5$:1 for $\alpha=1.5$ between
the hydrogen-burning limit and 0.01$M_\odot$. The ratio is $\sim2:1$, in the
same sense, for $\alpha=1.3$ at low masses (R99). 
However, if $\alpha=1.0$, then sub-solar
mass stars outnumber brown dwarfs by $\sim5:4$, while the ratio rises to $\sim7:1$ 
for $\alpha=0.5$. Converting to mass density, 
brown dwarfs add 40\%, 20\%, 7\% and 1\% to the local stellar mass density 
for $\alpha = 1.5, 1.3, 1.0$ and 0.5, respectively.

Statistically, the most significant result from the present analysis is that the only 
means of matching a steep, Salpeter-like ($\alpha\ge$2) mass function to
the present observations is by depleting the
immediate Solar Neighbourhood of all brown dwarfs younger than 10$^8$ years. As we 
noted above, the presence of active, $<10^8$-year old  G dwarfs within that same
volume argues against this extreme hypothesis. 
Thus, all current observational analyses indicate that it is extremely unlikely 
that brown dwarfs contribute substantially to the local mass density.

\section {Rotation and Activity}

Chromospheric activity in solar-type stars has long known to be well correlated
with the stellar rotational velocity. That correlation is due to the presence
of a magnetic $\alpha \Omega$ (shell) dynamo, generated by a toroidal field
located at the boundary between the radiative core and the convective envelope.
Traditionally, this paradigm has been
extended to M dwarfs, even though those stars are known to become
fully convective at spectral type $\approx$M4, Hawley {\sl et al.} (2000) outline an 
alternative model, where activity in mid- and late-type M dwarfs is driven largely
by a turbulent dynamo (Durney {\sl et al.}, 1993) within the convection
zone. The overall level of activity decreases significantly in dwarfs 
later than M9, with only a small number of L dwarfs showing 
detectable H$\alpha$ emission (G2000). This might 
reflect either decreased efficiency of the turbulent dynamo,  or the formation of 
radiative zones, which inhibit the emergence of magnetic flux, or increased
resistivity in the predominantly neutral atmospheres (Mohanty \& Basri, 2002).

If magnetic activity in late-type M dwarfs is powered by turbulence, one expects little
direct correlation between the level of activity and rotation. Hawley {\sl  et al.}
(2000) analysed the relatively scarce data available at that time, and found no evidence
for a significant correlation amongst late-type M dwarfs. This conclusion is generally
confirmed by Basri (2001) for a sample which includes 26 ultracool M dwarfs, ten of which
are in the present sample. The additional dwarfs observed here more than double the
number of ultracool M dwarfs with known rotational velocities, allowing us to
revisit this issue. 

Chromospheric activity in late-type dwarfs is generally gauged by measuring the
strength of H$\alpha$ emission. Table 2 lists equivalent width measurements for
all of the dwarfs in the present sample. However, those data should not be used directly 
to characterise activity: the equivalent width of an emission line depends on the
contrast with respect to the local continuum, rather than the absolute line flux.
As one moves down the M dwarf sequence, the continuum flux emitted at 6560\AA\
decreases, both in absolute terms and as a fraction of the bolometric flux. Thus, 
a 5\AA\ equivalent width H$\alpha$ line in an M3 dwarf represents significantly more chromospheric
flux than 5\AA\ emission in an ultracool M9. Reid, Hawley \& Mateo (1995a; hereinafter, RHM) originally
suggested that the appropriate method of dealing with this issue is, following
X-ray astronomy, computation of the normalised flux ratio, 
${F_\alpha \over F_{bol}} \ \equiv {L_\alpha \over L_{bol}}$,
the fraction of the luminosity emitted in H$\alpha$ emission.

In studying this issue, we have combined data from our own observations and
Basri's (2001) analysis, adopting our measurements for objects in common.
All of the dwarfs in our sample have low resolution spectroscopy,
allowing direct determination of the continuum flux at 6560\AA, F$_C$, and hence conversion
of H$\alpha$ equivalent widths to flux measurements. Similar data exist for
approximately half of the Basri sample; the remaining stars have either direct R-band
photometry or (V-I) and/or (I-J) colour measurements, from which we can estimate (R-I) and
hence R. Given F$_R$, we estimate F$_C$ using the empirical relation derived by
RHM, F$_C$=1.45F$_R$. 

All of the dwarfs in our sample and most of the Basri dwarfs have JHK/K$_S$
photometry. As discussed above, there is little variation in the J-band bolometric
corrections for dwarfs with spectral types between $\approx$M6 and L5; thus, for
dwarfs with near-infrared data, we adopt
\begin{displaymath}
m_{bol} \ = \ J \ + \ 2.1
\end{displaymath}
The remaining dwarfs have I-band photometry. I-band bolometric corrections are small
($<0.5$ mag.) for ultracool dwarfs, and we estimate BC$_I$
from the relations given in Reid \& Hawley (2000).

The derived flux ratios are listed in Table 2 with the equivalent width measurements. 
Figure 15a plots these results as a function of
$v \sin{i}$, differentiating amongst the different spectral types. There is no evidence 
for a significant correlation, either within a given spectral type or for the sample
as a whole. Nor is there evidence for a strong correlation between spectral type and 
rotation amongst the ultracool M dwarfs in our sample (Figure 15b). If we define
fast rotators as dwarfs with $v \sin{i} > 20$ kms$^{-1}$, the relative number of fast and
slow rotators is statistically identical at M7 (3/12), M8 (4/13) and M9 (1/7). There is 
marginal evidence that the lower envelope in $v \sin{i}$ increases to later spectral types:
three of the seven M9/M9.5 dwarfs (43\%) have $v \sin{i} < 10$ kms$^{-1}$ as compared with 11/15
(74\%) of the M7/M7.5 dwarfs. 
As discussed by Basri (2001), all L dwarfs which have been observed at sufficient spectroscopic
resolution to detect rotation have $v \sin{i} \ge 10$ kms$^{-1}$.

Turning to chromospheric activity in the present sample, we can consider three issues:
the mean level of activity, the dispersion in activity and the prevalence of substantial
flares. All three issues are discussed extensively by G2000, based on
lower-resolution spectroscopy of 60 late-M and L dwarfs. Our echelle observations resolve
weaker H$\alpha$ emission than was possible in that analysis, and therefore extend
coverage to lower activity levels, but the overall conclusions are unchanged.

Figure 15c plots activity, log($F_\alpha \over F_{bol}$), as a function of spectral
type for ultracool dwarfs from both the present sample and dwarfs from Basri (2001). The mean level
of activity amongst field M0 to M6 dwarfs with H$\alpha$ emission 
is well defined at log(${F_\alpha \over F_{bol}}) = -3.9$
(PMSU2, G2000). As Figure 15c shows, fewer than
10\% of the ultracool dwarfs in our sample reach this level, with several dwarfs
falling almost two orders of magnitude below the line. Our data reproduce the
trend of decreasing activity with later spectral type which is illustrated more
emphatically by the larger sample in G2000. Figure 15c also suggests that
the additional dwarfs from Basri's sample may exhibit lower activity at a given spectral type
than the average from our sample. Many of those stars are
 proper-motion, rather than photometrically,  selected.
One might conjecture that the lower activity amongst
the latter stars indicates a higher average age.

Finally, as mentioned in \S2.3, two of the dwarfs in our sample, 2M0320+18 (LP 412-31) and 
2M0350+18, have significantly stronger H$\alpha$ emission in our observations than in
prior measurements. Our previous observation of LP 412-31 indicates a high level
of activity, with a measured equivalent width of 40\AA\ (G2000); however, no H$\alpha$
emission was detected in the same authors' low-resolution spectrum of 2M0350+18, consistent with
an equivalent width of less than 4\AA. The inactivity of the latter star is particularly
unusual, given that over 90\% of the M8 and M9 dwarfs observed by G2000 have detectable
emission (2M0350+18 is classed as M9 by G2000). 
Our current observations indicate increased activity
by factors of two and at least 10 respectively. We note that the relative strength of the H$\alpha$
and He I 6678\AA\ emission lines are comparable in spectra taken during both 
 active and quiescent phases, suggesting  little variation in the plasma temperature.
There is also no evidence in the 2M0350+18 spectrum for any continuum distortion, such as was
seen during the 2M0149 superflare (Liebert {\sl et al.}, 1999). 

Ultracool dwarfs have been known to exhibit this type of behaviour since Herbig's (1956)
spectroscopy of VB 10, and \pedant \& Ardila (2001) summarise more recent flare observations.
Our detection of two events amongst a sample of thirty-nine dwarfs, all with similar observing
times, indicates an overall duty cycle of $\approx$5\%, broadly consistent with previous
estimates (Reid {\sl  et al.}, 1999b; G2000; \pedant \& Ardila, 2001).  
However, there are qualitative differences between the two events detected here. RHM
 originally pointed out that variations of up to a factor of two in
equivalent width were not uncommmon in repeated observations of active M dwarfs in both
the Hyades and the field: the 2M0320+18 event detected here lies in this range of variation, 
as does the \pedant (1999) observation of VB8. On the other hand, the flares observed in
2M0350+18 (this paper), BRI0021 (Reid {\sl  et al.}, 1999b), LHS 2065 (\pedant \& Ardila, 2000), 
2M0149+29 (Liebert {\sl et al.}, 1999),  LHS 2397a (Bessell, 1991) and even VB 10 (Herbig, 1956)
are significantly more energetic, with the H$\alpha$ flux enhanced by factors from 20 to $>100$.
Spectroscopic monitoring of a sample of ultracool dwarfs would provide interesting information
on the relative frequency of these events as a function of their intensity.

\section { Distances and Kinematics}

The primary aim in compiling a photometrically-selected sample of ultracool dwarfs is
avoiding bias toward high-velocity, old M dwarfs. As a corollary, the 
current sample should provide reliable statistics on the solar motion and velocity 
dispersions of these late-type dwarfs. Those data, in turn, offer a means of probing
the likely age distribution of the sample. Several previous analyses have suggested
that ultracool dwarfs are younger, on average, than earlier-type M dwarfs (Hawkins \& Bessell, 1988;
Kirkpatrick {\sl et al.},
1994; Reid, Tinney \& Mould, 1994). The most extensive previous kinematic
study, by Tinney \& Reid (1999), found no statistical evidence 
that ultracool dwarfs were drawn from a different velocity distribution, but
that analysis includes only 13 photometrically-selected dwarfs. We can
re-examine this issue with our larger sample. 

\subsection {Distance determination}

Our spectra provide direct measurement of radial
velocities, and proper motions are available from the literature for most of the sample. 
All of the remaining targets in the photometrically-selected 
sample are easily visible on UKST and/or POSS II photographic sky survey plates;
indeed, most are clearly detected on the POSS I E plates. As
discussed by G2000, those data, combined with 2MASS astrometry,
provide sufficient baseline for proper motion measurements. All of the proper
motion measurements are listed in Table 5. 

Deriving transverse velocities, however, requires both proper motion
measurements and distance estimates. 
Trigonometric parallaxes have been measured for only 8 of the 37 systems in
our photometrically-defined sample. For the remaining dwarfs,  we must
resort to photometric parallax. All of the dwarfs have JHK$_S$ photometry, so we
can follow G2000 and estimate M$_K$ from the the (J-K$_S$) colours using the
following relation, 
\begin{displaymath}
M_K \ = \  7.593 \ + \ 2.25 \times (J-K_S), \qquad \sigma_{rms} = 0.36 \ {\rm mag.}
\end{displaymath}
However, there is a potential complication, in that the 
current sample was colour-selected, with the requirement (J-K$_S)>0.95$. This
opens the possibility for bias, since the dispersion in the (M$_J$, (J-K$_S$)
main-sequence at higher luminosities is sufficient to allow redder, higher luminosity
stars to scatter into the sample (Figure 1). This is highlighted by the 
location of the M6 dwarfs in Figure 2.

Fortunately, most of the dwarfs in the photometrically-defined sample also have I-band
photometry, either based on direct measurements or synthesised from our low-resolution
spectroscopy. We can use those colours to estimate distances based on the photometric
parallax calibration derived by Reid \& Cruz (2002),
\begin{eqnarray*}
M_I & = & 16.491 - 16.499 (I-J) +14.003 (I-J)^2 - 4.717 (I-J)^3+ 0.697 (I-J)^4 \\
& &- 0.0330 (I-J)^5, \quad  1.65 \le (I-J) < 4.0, \ \sigma = 0.31 {\rm \ mag., \ 37 \ stars}
\end{eqnarray*}
Figure 16 compares distance moduli derived from the limited trigonometric parallax data and
from the (J-K$_S$) and (I-J) colour indices. The upper panel shows that there is no evidence for
a systematic difference between the trigonometric and photometric indicators, which is not
surprising, since several of these dwarfs were included in calibrating the latter. One star
deserves special comment: the (I-J) and (J-K$_S$) photometric parallaxes for RG0050.5 are in
excellent agreement, indicating a distance of 30.5 parsecs, while trigonometric measurements
give 22.2 parsecs. Thus, this star is bluer than expected, and lies $\sim0.7$ magnitudes below 
the mean relation for M8 dwarfs.

There is, however, a clear systematic trend when we compare the photometric distance indicators, with
the (J-K$_S$) calibration tending to derive lower distances (by $\sim20\%$), and fainter absolute magnitudes.
We find
\begin{displaymath}
\delta (m-M) \ = \ (m-M)_{I-J} \ - \ (m-M)_{J-K} \ = \ 0.35\pm0.11, \quad {\rm 29 \ stars}
\end{displaymath}
Given this comparison, we use (I-J) in preference to (J-K$_S$) in computing the distances
listed in Table 5. The uncertainties therein are based on the 
 dispersions in the calibrating colour-magnitude relations.

\subsection {Kinematics of ultracool dwarfs}

We have used our distances estimates for each dwarf to derive V$_\alpha$ and V$_\delta$
from the proper motion measurements, and combined those data with
the radial velocity to derive the Galactic space motions  listed
in Table 5. The velocity distribution is shown in Figure 17. Following the
convention of the Catalogue of Nearby Stars (Gliese, 1969), the (U, V, W) motions are
defined as a right-handed system (U positive towards the Galactic Centre).

If we consider the sample as a whole (including all 37 systems which meet the
apparent magnitude selection criteria), the mean motions and dispersions are
\begin{equation}
({\rm U, V, W}; \ \sigma_U, \sigma_V, \sigma_W) \ = \ (-23.9, -14.6, -9.5; \ 31.1, 16.2, 16.6) \ 
{\rm km s^{-1}}
\end{equation}
Excluding 2M0746+20 and the two M6/M6.5 dwarfs results in minimal changes,
\begin{equation}
({\rm U, V, W}; \ \sigma_U, \sigma_V, \sigma_W) \ = \ (-23.7, -14.2, -9.3; \ 32.1, 17.1, 16.9) \ 
{\rm km s^{-1}}
\end{equation}
The two most extreme outliers in the velocity distribution are 2M0109+29 (M9.5) and 2M1403+30
(M8.5), Both are identified in Figure 17, and in both cases the significant uncertainties
reflect the extent to which the velocity estimates rest on the transverse motions.
Excluding those two dwarfs from the analysis gives 
\begin{equation}
({\rm U, V, W}; \ \sigma_U, \sigma_V, \sigma_W) \ = \ (-18.5, -12.1, -9.8; \ 26.7, 15.9, 15.9) \ 
{\rm km s^{-1}}
\end{equation}
based on data for 31 ultracool dwarfs (M7 {--} M9.5). 

We have used different symbols in figure 17 to identify dwarfs of different spectral type.
Inspection of those diagrams shows limited evidence for significant variation in
kinematics, and that conclusion is broadly confirmed by statistical analysis.
Given the small sample size, we
can only subdivide to a limited extent. However, we have computed mean motions for
the fourteen M7/M7.5 dwarfs, finding 
\begin{equation}
({\rm U, V, W}; \ \sigma_U, \sigma_V, \sigma_W) \ = \ (-20, -13, -10; \ 32, 21, 14) \ 
{\rm km s^{-1}}
\end{equation}
and for the fourteen M8/M8.5 dwarfs, 
\begin{equation}
({\rm U, V, W}; \ \sigma_U, \sigma_V, \sigma_W) \ = \ (-18, -16, -7; \ 27, 17, 15) \ 
{\rm km s^{-1}}
\end{equation}
These two sets of results are statistically indistinguishable.
We therefore take the kinematics listed in equation 2, derived from 34 M7 to M9.5 dwarfs, as
characteristic of the ultracool dwarf sample. 

We can compare these results against similar analyses of higher-mass M dwarfs. Hawley {\sl et al.}
(PMSU2) have calculated space motions for a volume complete sample of early- and
mid-type M dwarfs in the Solar Neighbourhood, making an explicit division between stars 
with and without H$\alpha$ emission. Combining data for all M dwarfs, they derive
\begin{equation}
({\rm U, V, W}; \ \sigma_U, \sigma_V, \sigma_W) \ = \ (-10, -21, -8; \ 38, 26, 21) \ 
{\rm km s^{-1}}
\end{equation}
For non emission-line dM dwarfs, they derive
\begin{equation}
({\rm U, V, W}; \ \sigma_U, \sigma_V, \sigma_W) \ = \ (-9, -23, -8; \ 41, 27, 21) \ 
{\rm km s^{-1}}
\end{equation}
The velocity dispersions are significantly higher than our measurements for ultracool
dwarfs. In contrast, for the emission line dwarfs, they find
\begin{equation}
({\rm U, V, W}; \ \sigma_U, \sigma_V, \sigma_W) \ = \ (-12, -13, -8; \ 27, 20, 15) \ 
{\rm km s^{-1}}
\end{equation}
values much closer to the analysis of the ultracool sample. 

Dahn {\sl et al.} (2002) have recently undertaken a similar comparison, 
using tangential velocities for a sample of 28 ultracool dwarfs, including 
eight with spectral type M8 to M9.5, 17 L dwarfs and 3 T dwarfs.  
There are only two stars in common with the present sample. However, they
also find that the kinematics of this L-dwarf dominated sample are
very similar to those of the PMSU dMe stars. 

Velocity dispersions provide only a one-dimensional parameterisation of the
velocity distribution; in particular, they provide no indication of how well
the overall distribution matches a Gaussian. Probability plots (Lutz \& Upgren, 1980)
serve that function: a Gaussian distribution gives a straight line if one
plots the cumulative distribution in units of the measured standard deviation, $\sigma$.
Figure 18 plots such data for the U, V and W distributions defined by the 34 systems in
our ultracool dwarf sample (M7-M9.5), 73 dMe systems and 355 dM systems. The latter
are volume-limited samples with $\delta > -30^o$ and  
absolute magnitudes in the range $7.0 \le {\rm M_V} < 15.0$ (M0 to M6, 
see Reid, Hawley \& Gizis, 1995;  PMSU1). There is clearly much closer agreement between the 
distributions outlined by the ultracool dwarfs and the dMe dwarfs. 

We can compare these  distributions quantitatively using the Kolmogorov-Smirnov 
test. That comparison shows that there is a probability of more than 10\% that the
ultracool dwarfs and dMe stars are drawn from the same kinematic population: the two
datasets are statistically indistinguishable. However, while the same
holds for a comparison between the W-velocity dispersion of the ultracool
dwarfs and the dM sample, there is a probability of less than 5\% that the
U and V distributions are drawn from the same parent population. Comparing the
ultracool dwarf sample against the complete PMSU1 dataset (dM+dMe) reveals
inconsistencies at the same level, as one might expect given the predominance of
the non-emission line M dwarfs.  

Can the relatively low space motions of the ultracool dwarfs be attributed to
a systematic error in our distance scale? The present sample were identified
using photometric criteria (location in the (K$_S$, (J-K$_s$)) plane), and 
most have distance estimates derived from photometric parallax. Our use of
(I-J) rather than (J-K$_S$) should compensate to a large extent for stars
introduced into the sample through scatter in colour (as illustrated in
Figure 16). However, the dispersion in absolute magnitude about the mean
main sequence can also lead the inclusion of higher luminosity, more
distant stars, and a statistical underestimation of both the average distances
and tangential velocities. 

We can estimate the likely extent of this effect using classical
Malmquist bias: if we select stars from a uniform density distribution, the
mean absolute magnitude of the sample, $\bar M$, is given by
\begin{equation}
\bar {M} \ = \ M_0 \ - \ 1.38 \sigma^2
\end{equation}
where M$_0$ is the mean absolute magnitude of a volume limited sample, and
$\sigma$ the rms uncertainty associated with the absolute magnitude calibration.
The (M$_I$, (I-J)) relation has a dispersion of 0.31 magnitudes; the (M$_K$, (J-K$_S$))
calibration given by G2000 has a dispersion of 0.36 magnitudes. 
Adopting the latter value gives a statistical offset of 0.18 magnitudes in equation 9,
or an average underestimate of $\sim9\%$ in distance. We have re-calculated the 
kinematics for the 34 M7 to M9.5 dwarfs, increasing {\sl all} of the distances (including
trigonometric parallax measurements) by 10\%. The resultant kinematics are
\begin{equation}
({\rm U, V, W}; \ \sigma_U, \sigma_V, \sigma_W) \ = \ (-24.7, -15.4, -9.7; \ 34.2, 18.6, 17.5) \ 
{\rm km s^{-1}}
\end{equation}
These still fall short of the M dwarf kinematics given in equation (6). 

Our conclusion is that the ultracool dwarfs in the
present sample have kinematics which are statistically very similar to
those of Solar Neighbourhood dMe dwarfs. The velocity dispersions are
significantly lower than those of the
non-emission line M dwarfs in the local old disk population, but
intriguingly similar to the mean kinematics of the L-dwarf dominated sample
analysed by Dahn {\sl et al.} (2002). 

\subsection {Discussion}

An obvious candidate for the observed discrepancy between the kinematics of ultracool
M dwarfs and the average kinematics of earlier-type M dwarfs is a difference in 
mean age between the two samples. 
H$\alpha$ emission declines with age (at a mass-dependent rate), so the 
dMe sample has a younger mean age than the dM sample, and, as a consequence,
cooler kinematics. If we assume a constant star formation rate over
the history of the disk, the relative numbers in the dM and dMe samples suggest
that the latter stars are drawn from the most recent 15-20\% of disk star
formation, or ages up to 2 Gyrs for a canonical 10-Gyr disk. Matching the
velocity dispersions of the dMe sample against Jahrei{\ss} \& Wielen's
(1983) calibration gives a slightly older age, $\tau = 3\pm1$ Gyrs. The 
similar kinematics shown by the ultracool dwarfs suggests that they also
have ages of  $\tau <  2$ to 3 Gyrs.

Is an age difference a physically reasonable explanation for the different
kinematics of dM and ultracool dwarfs? In the case of the Dahn {\sl et al.}
analysis, that hypothesis is not unreasonable, since a significant fraction of
the L and T dwarfs in their sample are likely to be substellar mass
brown dwarfs. Given the apparent similarity in motions, it is tempting
to ascribe the results of our current analysis to the same underlying cause.

However, as Figures 8, 13 and 14 show, we expect M-type ultracool dwarfs to include a 
mixture of very low-mass stars and brown dwarfs, with the former dominating
the sample. All M dwarfs have main-sequence lifetimes well in excess of a
Hubble time, so both the ultracool dwarfs and the PMSU1 dM+dMe sample should 
include representation from the oldest stars in the Galactic disk. Our simulations
predict an average age of $\sim4.5$ Gyrs for $\tau_{max} = 10$ Gyrs and
a uniform birthrate. Given a near-exponential birthrate, such as that favoured by
cosmological studies (Madau {\sl et al}, 1998), the average age rises to $\sim7$ Gyrs.
In either case, the observed velocity dispersions of the ultracool dwarfs are
significantly lower than expected for a sample dominated by hydrogen-burning dwarfs.

If we accept that kinematics are reliably correlated with age, and that 
our observations provide a fair sample of the ultracool dwarf population, 
then there appear to be two possible explanations for these observational results:
\begin{enumerate}
\item the Solar Neighbourhood is deficient in very low-mass ($M < 0.09 M_\odot$)
dwarfs older than $\sim 4$ Gyrs. This might reflect sampling of the local disk
population, or a significant increase in star formation within the last 5 Gyrs (that is,
opposite to the cosmological trend favoured by Madau {\sl et al.}, 1998). Such
a change seems unlikely, although most age indicators, such as chromospheric
activity, become less reliable at ages exceeding $\sim2$ Gyrs.
In any event, this conclusion implies a steepening in the inferred mass function at
low masses, since analyses assume that the observed numbers of low-mass stars
reflect formation throughout the entire lifetime of the disk. 
\item low-mass dwarfs have lifetimes of $<4$ Gyrs as M7 to M9.5 dwarfs. This 
effectively requires that the majority of ultracool dwarfs are
high-mass brown dwarfs or transition objects (the return of M dwarfs as
brown dwarfs in masquerade). That hypothesis, in turn, requires that
either the spectral-type/temperature scale adopted here is incorrect, in the
sense that late-type dwarfs lie at cooler temperatures (i.e an adjustment 
in the opposite sense to that favoured by Basri {\sl et al.}, 2001), or 
that both sets of evolutionary models plotted in Figure 8 are incorrect
in their location of the hydrogen-burning limit.
\end{enumerate}
If the latter option is correct, and brown dwarfs dominate the
ultracool sample, then we also require a flat or decreasing decreasing mass function
($\alpha < 0.5$) to accommodate the observed lithium fraction.

The referee has suggested that significant revision in the evolutionary
models is unlikely given the good agreement between those models and observations 
of the binary brown dwarf, 
Gl 569Bab (Lane {\sl et al.}, 2001). We note, however, that the agreement hinges
on the age of 300 Myrs associated with the system. That age estimate
derives partly from the observed levels of chromospheric and coronal activity 
of the primary, Gl 569A, which
Lane {\sl et al.} argue are consistent with an age between 0.2 and 1 Gyr, and
partly from an hypothesised association with the Ursa Major moving group
(Kenworthy {\sl et al.}, 2001). Analysing data for higher-mass (M$_V < 7$)
members, Soderblom \& Mayor (1993) use isochrone fitting
to derive an age of 300 to 400 Myrs for the latter moving group.

Closer inspection of the data, however, suggest an older age. 
First, the observed radial velocity of $-6.9\pm1.0$ kms$^{-1}$
(Gizis {\sl et al.}, 2002) is significantly different from 
the predicted value of -0.6 kms$^{-1}$ (Madsen {\sl et al.}, 2002).
This peculiar velocity corresponds to a drift of $\sim6.5$ pc Myr.$^{-1}$
relative to the centroid of the moving group,  
or more than 2 kpc. over its lifetime. This suggests that membership is unlikely.
Second, the intrinsic properties of Gl 569A
favour an older age: Gl 569A has a level of
X-ray activity ($L_X \over L_{bol}$) matching the average level of Hyades stars 
of similar luminosity and spectral type (age $\sim 600$ Myrs);
the chromospheric activity ($L_\alpha \over L_{bol}$) is lower than most
of Hyades dwarfs; and the star lies in the middle of the (M$_V$, (V-I))
main-sequence defined by nearby field dwarfs. All of these properties are
more consistent with an age between 0.6 and 1 Gyr. than the $\sim 300$ Hyrs
favoured by Lane {\sl et al.}  Adopting the older ages leads to 
evolutionary masses of 0.07 to 0.09 M$_\odot$ for both components
of Gl 569Bab, and a predicted total mass 20 to 50\% higher than the dynamical measurement
of $0.123^{+0.027}_{-0.022} M_\odot$.

It remains possible that the sample of ultracool dwarfs discussed in this
paper is biased in some respect. We
are currently undertaking a large-scale survey which aims to identify all
ultracool dwarfs within 20 parsecs of the Sun (Cruz {\sl et al.}, in prep.). 
High resolution observations of those dwarfs, coupled with trigonometric 
parallax measurements, should help solidify our understanding of this issue.

\section {Conclusions}

We have presented high-resolution optical spectroscopy of 39 ultracool dwarfs. Our
prime goal is the detection of absorption due to Li I at 6708\AA, 
an unequivocal indication of substellar mass at these temperatures. 
Two dwarfs exhibit significant absorption:
LP 944-20, identified previously as a brown dwarf by Tinney (1998), and 2MASSW J0335020+234235.
Both are included in a 33-dwarf photometrically-selected sub-sample, 
spanning spectral types M9.5 to M7. We have compared the observed fraction of lithium-rich
dwarfs, $F_{Li}$, against predictions from simulations based on theoretical tracks
calculated for low-mass dwarfs by both Burrows {\sl et al.} (1997) and Baraffe {\sl et al.} (1998),
employing a range of power-law mass functions and assuming a constant birthrate. 
Our models show that $F_{Li}$ depends strongly on the proportion of young ($\tau < 10^8$ years)
dwarfs in the Solar Neighbourhood, a result stemming from the correlation between mass and cooling time,
and reflecting the temperature range spanned by ultracool dwarfs.
Moreover, the two sets of theoretical trackss predict lithium fractions which differ by over 50\%.
Nonetheless, it seems clear that the observed values of $F_{Li}=6\pm4$\% for M7 to M9.5
dwarfs, and $10\pm7$\% for M8 to M9.5 dwarfs, are only consistent with near-Salpeter 
mass functions ($\alpha > 2$) if the Solar Neighbourhood is completely deficient of
brown dwarfs with ages $\tau < 10^8$ years. This seems unlikely, given the observed
distribution of chromospheric activity amongst G dwarfs in the same volume. 

We have also considered the rotational properties of our sample of ultracool dwarfs,
and the correlation between rotation and chromospheric activity. Our observations
confirm Hawley {\sl et al.'s} (2000) conclusion that there is no significant
correlation between the latter two parameters for late-type M dwarfs. The relative number
of fast ($v \sin{i} > 20$ kms$^{-1}$) and slow rotators is invariant with spectral type, 
although there is marginal evidence that the average rotational velocity of M9/9.5 dwarfs 
is higher than amongst earlier types. The range of chromospheric activity exhibited
in our observations is consistent with previous studies (G2000; Basri, 2001), 
with the typical H$\alpha$ flux lying below the mean level for earlier-type M dwarfs.
Two of the 39 dwarfs have H$\alpha$ fluxes in our observations
which are significantly higher than previous
measurements. The inferred duty cycle of 5\% for flare outbursts is consistent with
previous estimates for late-type dwarfs.

Finally, we have combined our radial velocity measurements with proper motion data and
distance estimates to derive space motions for the ultracool dwarfs in our sample. 
The mean kinematics are almost identical to those derived by Hawley {\sl et al.} (1996)
for a volume-limited sample of M0 to M6 emission-line dwarfs. This result is surprising, given
that the ultracool dwarf sample is expected to be dominated by long-lived, 
very low-mass stars, with an average age of 4 to 5 Gyrs (for a disk age of 10 Gyrs). 
This discrepancy raises serious questions concerning both the existence of possible biases
in the sampling of the Galactic Disk stellar populations offered in the immediate
Solar Neighbourhood, and the reliability of evolutionary models for low mass stars and brown
dwarfs. Detailed analysis of high-resolution spectroscopic observations of a larger sample
of ultracool dwarfs is required before those questions can be answered in a satisfactory manner.

\acknowledgements 
We thank the referee for useful comments.
The initial research for this project
was supported partially by a Core Science grant from the 2MASS
project.  This
publication makes use of data products from the Two Micron All Sky Survey,
which is a joint project of the University of Massachusetts and the Infrared
Processing and Analysis Center/California Institute of Technology, funded by
the National Aeronautics and Space Administration and the National Science
Foundation.

\begin{table}
\caption{Target stars}
\begin{center}\scriptsize
\begin{tabular}{lccccccccc}\hline
2MASS & Other name & I &Ref. & (J-H) & (H-K$_S$) & K$_S$ & Sp. type  & Src.  &Epoch \\
\hline\hline 
I J0052546-270559 & RG0050.5-2722 & 16.82 &A& 0.71  & 0.42 & 12.56 & M8 & 1, 2&12/99 \\
W J0109216+294925 & & 16.01 &B& 0.73 & 0.49 & 11.70 & M9.5 &3 & 12/99 \\
W J0140026+270150 & & 15.64 &B& 0.69 & 0.38 & 11.44 & M9.5 &3 &12/99 \\
W J0149089+295613 & & 16.73&B& 0.86 & 0.56 & 11.99 & M8.5 &3&12/99 \\
I J0253202+271333 & & 14.98&B&0.67 & 0.37 & 11.45 & M8 &3 &12/99 \\
W J0320597+185423 & LP 412-31 &14.70&C& 0.70 & 0.47 & 10.57 & M8 &2&12/99 \\
I J0330050+240528 &  & 14.84&B&0.61 & 0.39 & 11.36 & M7 &3& 12/99 \\
I J0335020+234235 &  & 14.63&B&0.61 & 0.39 & 11.26 & M8.5 &3& 12/99 \\
W J0339352-352544 & LP 944-20 &14.16&A& 0.73 & 0.50 & 9.52 & M9.5 & 4 & \\
W J0339528+245726 &  & & & 0.67 & 0.46 & 11.73 & M8   &5& 3/99 \\
W J0350573+181806 &  & 15.59&B&0.73 & 0.46 & 11.76 & M8 &3& 12/99  \\
W J0429028+133759 & LP 475-855 &15.39&B& 0.69 & 0.34 & 11.64 & M7 &6& 3/99 \\
W J0746425+200032 &  &15.11&C& 0.74 & 0.51 & 10.49 & L0.5 &7& 3/99 \\
W J0810586+142039 &  &15.35&B& 0.67 & 0.43 & 11.61 & M9 &3& 3/99 \\
W J0818580+233352 &  &14.40&B& 0.64 & 0.37 & 11.31 & M7 &3& 12/99 \\
W J0853361-032931 & LHS 2065&14.44&D& 0.75  &0.51 & 9.98 & M9   &2& 3/99 \\
W J0925348+170441 &  &14.60&B& 0.61 & 0.39 & 11.60 & M7 &3& 12/99 \\
W J0952219-192431 &  &14.49&B& 0.60 & 0.43 & 10.85 & M7 &3& 12/99 \\
W J1016347+275150 & LHS 2243 &14.42&B& 0.66 & 0.34 & 10.95 & M8 &2& 3/99 \\
W J1047127+402644 & LP 213-67 &13.17&B& 0.64 & 0.41 & 10.40 & M6.5 &8& 6/99 \\
W J1047138+402649 & LP 213-68 &14.75&B& 0.64 & 0.43 & 11.28 & M8 &8& 6/99 \\
W J1200329+204851 &  &15.14&B& 0.60 & 0.43 & 11.82 & M7 &3& 12/99 \\
W J1224522-123835 & BRI 1222-1221 &15.74&A& 0.73 & 0.46 & 11.37 & M9 &2& 6/99 \\
P J1242464+292619 & & & & 0.65 & 0.55 & 13.23 & M8 & 5 & 6/99 \\
W J1246517+314811 & LHS 2632 &14.70&B& 0.67 & 0.36 & 11.23 & M7.5 &2& 3/99 \\
W J1253124+403404 & LHS 2645&14.70&B& 0.62 & 0.38 & 11.17 & M7.5  &2& 3/99 \\
P J1309219-233035 &  &14.96&B&  0.68& 0.42 & 10.67 & M8 &5& 3/99 \\
W J1336504+475131 &  & & & 0.58 & 0.43 & 11.63 & M7 &3& 12/99 \\
W J1403223+300755 &  &  & & 0.68 & 0.38 & 11.63 & M8.5 &3& 12/99 \\
I J1501081+225001 & TVLM513-46546 &15.21&C& 0.68 & 0.40 & 10.72 & M8.5 &9& 3/99 \\
W J1504162-235556 &  & & &  0.64  & 0.36 & 11.03  & M7.5  &3 & 6/99 \\
P J1524248+292535 &  & & & 0.68 & 0.39 & 10.15 & M7.5  &5& 3/99 \\
W J1527194+413047 &  & & & 0.63 & 0.38  & 11.47  & M7.5 &3 & 6/99 \\
W J1550382+304108 &  &  & &0.58 & 0.49 & 11.92 & M7.5&3& 6/99 \\
W J1714523+301941 &  &  & &0.66 & 0.39 & 11.89 & M6.5 &3& 6/99 \\
W J2206228-204705 &  & 15.05&B&0.68 & 0.40 & 11.35 & M8 &3& 6/99 \\
W J2233478+354747 &  &  & &0.64 & 0.42 & 10.88 & M6 &3& 6/99 \\
W J2234138+235956 &  & 16.42&B&0.81 & 0.52 & 11.81 & M9.5 &3& 6/99 \\
W J2235490+184029 &  & 14.85&B&0.63 & 0.50 & 11.33 & M7 &3& 6/99 \\
\tableline      
\end{tabular}
\end{center}
References: photometry - 
A: Tinney, 1996; B: Flux-calibrated spectroscopy;
C: Dahn {\sl et al.}, 2000; D: Leggett, 1992. \\
Spectral-type  source - 
1. Reid \& Gilmore, 1981; 
2. Kirkpatrick, Henry  \& Simons, 1995; \\
3. Gizis {\sl  et al.}, 2000a;  4. Tinney, 1998; 
5. Kirkpatrick {\sl  et al.}, 1997; \\
6. Leggett, Harris \& Dahn, 1994; 
7. Reid {\sl  at al.}, 2000;
8. Gizis {\sl  et al.}, 2000b; \\
9. Tinney {\sl et al.}, 1993.
\end{table}

\clearpage

\begin{table}
\caption{Lithium, activity, radial velocities and rotation}
\begin{center}\scriptsize
\begin{tabular}{lcccccccc}\hline
2MASS & lithium  & H$\alpha$ & He I& log$F_\alpha \over F_{bol}$ & V$_{\alpha}$&V$_{CCF}$ & v sin(i)\\
   &  \AA & \AA & \AA & km s$^{-1}$ & km s$^{-1}$& km s$^{-1}$ \\ 
\hline\hline 
2M0052-27  & $<0.1$ &  1.2&     & -5.06 & $15.9\pm3.0$ & $16.8\pm0.5$ & $<4$\\
2M0109+29   & $<0.1$ &  0.6&    & -5.57 & $39.2\pm2.0$ & $38.9\pm0.8$ & 7 \\
2M0140+27   & $<0.1$ &  3.0&    & -5.04 & $ 9.6\pm2.0$ & $ 8.2\pm0.35$ & 6.5\\
2M0149+29   & $<0.16$&  5.5&    & -4.83 & $27.9\pm2.0$ & $28.1\pm0.8$& 12 \\
2M0253+27   & $<0.1$ &  8.6&    & -4.51 &$39.3\pm2.0$ & $38.9\pm0.5$/$53.7\pm1.1$ & 5/7\\
2M0320+18   & $<0.1$ & 82.8& 4.3& -3.94 & $41.6\pm3.0$ & $44.7\pm0.6$& 8 \\
2M0330+24   & $<0.05$& 12.9&    & -4.29 & $39.3\pm2.0$ & $39.2\pm1.1$& 17 \\
2M0335+32   &   0.72 &  6.5&    & -4.63 &$ 9.6\pm2.0$ & $12.2\pm3.6$& 30\\
2M0339-35   & 0.53 &  1.2 &  & -4.82 &  & $10.0\pm2.0$ & 28 \\
2M0339+24   & $<0.15$&  9.9&    & -3.97 & $37.4\pm2.0$ & $34.9\pm 1.2$& 16\\
2M0350+18   & $<0.15$& 39.7& 2.5& -4.06 &  $ 5.8\pm2.0$ & $-14.3\pm0.4$ & 4 \\
2M0429+13   & $<0.1$ &  6.0& 0.2& -4.13 & $44.3\pm2.0$ & $44.3\pm1.1$ & 6\\
2M0746+20   & $<0.05$&  1.2&    & -5.24 & $49.2\pm3.0$ & $54.9\pm1.5$ & 24 \\
2M0810+14   & $<0.05$& 12.3&    & -4.48 & $27.8\pm2.0$ & $27.4\pm0.4$& 11\\
2M0818+23   & $<0.05$&  6.3&    & -4.41 & $33.4\pm2.0$ & $34.8\pm0.4$& $<4$ \\
2M0853-03   & $<0.05$& 12.6&    & -3.91 & $ 9.1\pm2.0$ & 8.7 & 11\\
2M0925+17   & $<0.1$ &  3.5&    & -4.47 & $11.4\pm2.0$ & $12.3\pm0.4$ & 21\\
2M0952-19   & $<0.1$ & 13.0& 0.3& -3.79 & $-15.0\pm2.0$ & $-25.9\pm0.5$/$-9.7\pm1.1$ & 5/7\\
2M1016+27   & $<0.1$ & 19.5& 1.1& -3.99 & $17.6\pm2.0$ & $17.1\pm0.8$& $<4$ \\
2M1047+40A  & $<0.05$&  4.5&    & -4.03 & $ 4.7\pm4.0$ & $5.1 \pm 3.5$ & 35\\
2M1047+40B  & $<0.1$ &  2.7&    & -4.73 & 17.5:: & $-1.1 \pm 3.6$& 35 \\
2M1200+20   & $<0.07$&  2.9&    &-4.64 & $-32.6\pm2.0$ & $-32.4\pm0.5$ & 6\\
2M1224-12   & $<0.07$& 21.3& 0.6& -3.59 & $-6.3\pm2.0$ & $-5.6\pm0.4$& 8 \\
2M1242+29   & 0.4: & 15.8 & & -4.13 & $-33.7\pm2.0$ & $-35.5\pm0.7$& $<4$\\
2M1246+31   & $<0.1$ &  0.6&    & -4.54 & $ 4.7\pm4.0$ & $6.7\pm0.7$& $<4$ \\
2M1253+40   & $<0.05$&  6.1&    & -3.87 &$ 2.6\pm2.0$ & $3.2\pm0.25$ & 6\\
2M1309-23   & $<0.05$  &  7.1&    & -4.76 & $18.3\pm2.0$ & $19.8\pm0.5$ & 7\\
2M1336+47   & $<0.13$&  3.2&    & -5.18 &$-13.6\pm2.0$ & $-3.1\pm7.5$ & 30 \\
2M1403+30   & $<0.05$ &  5.2&    & -5.10 &$-41.5\pm2.0$ & $-39.2\pm0.4$& 6 \\
2M1501+22   & $<0.05$&  3.5&    & -4.98 &$-2.4\pm3.0$ & $5\pm6$& $>40$ \\
2M1504-23   &$<0.07$ &  5.5&0.35& -4.97 & $-31.2\pm2.0$ & $-35.6\pm4.2$& 30  \\
2M1524+29   &$<0.05$ &  5.2&    & -4.47 &$-15.5\pm2.0$ & $-15.9\pm0.5$& 7  \\
2M1527+41   &$<0.1$  &  5.3&    & -4.33 &$-0.4\pm2.0$ &$3.0\pm1.8$& 7 \\
2M1550+30   &$<0.05$ & 16.9&0.6 & -4.30 &$-46.4\pm2.0$ & $-45.2\pm0.7$& 5 \\
2M1714+30   &$<0.1$  &  3.2&    & -4.96 &$-42.6\pm2.0$ & $-41.2\pm0.4$& $<4$\\
2M2206-20   &$<0.05$ &  5.1&    & -4.54 &$ 8.0\pm2.0$ & $16.3\pm2.7$& 22 \\
2M2233+35   &$<0.05$ &  5.3& 0.3& -4.66 & $-15.5\pm2.0$ & $-17.0\pm0.4$& 6\\
2M2234+23   &$<0.05$ & 21.5&0.25& -4.16 & $17.3\pm2.0$ & $17.2\pm0.65$& $<4$\\
2M2235+18   &$<0.08$ &  5.3&    &- 4.56 & $-16.3\pm2.0$ & $-16.0\pm0.6$& 9 \\
Gl 83.1 & \null  &  1.5&    & &$-31.3\pm2.5$ & -28.6 & $3.8\pm1.6$\\ 
Gl 412B & \null  &  9.5& 0.4& &$68.15\pm2.0$ & 68$$ & $7.0\pm1.7$ \\
\tableline      
\end{tabular}
\end{center}
Notes: \\
 2M0339-35 $\equiv$ LP 944-20:   V$_{CCF}$ from Tinney (1998), $v \sin{i}$ from Basri (2001); \\
 2M0853-03 $\equiv$ LHS 2065: V$_{CCF}$ from Tinney \& Reid (1999). \\ 
 Gl 83.1: V$_{CCF}$ from Marcy \& Benitz,  v$\sin{i}$ from  Delfosse {\sl  et al.} (1998); \\
 Gl 412B: V$_{CCF}$ and v$\sin{i}$ from Delfosse {\sl  et al.} (1998). \\
\end{table}

\clearpage

\begin{table}
\caption{Rotation \& Activity - comparison with previous observations}
\begin{center}\scriptsize
\begin{tabular}{lccccccc}\hline
     & Basri (2000) & & this paper & \\
Name & $v \sin{i}$ & H$\alpha$ & $v \sin{i}$ & H$\alpha$ \\
 & kms$^{-1}$ & \AA\ &  kms$^{-1}$ & \AA\ \\
\hline\hline 
LP 412-31 & 9.0 & 18.4 & 8 & 82.8 \\
LHS 2632 & $<3$ & 0.4 & $<4$ & 0.6 \\
LHS 2645 & 6.5 & 4.9 & 6 & 6.1 \\
RG0050.5 & $<5$ & 2.9 & $<4$ & 1.2 \\
LHS 2243 & $<5$ & 15.8 & $<4$ & 19.5 \\
2M1242+29 & 5.0 & 8.1 & $<4$ & 15.8 \\
TVLM513-46546 & 60 & 1.8 & $>40$ & 3.5 \\
LHS 2065 & 9 & 8.4 & 11 & 12.6 \\
BRI1222 & 2.0 & 9.7 & 8 & 21.3 \\
\tableline      
\end{tabular}
\end{center}
\end{table}
\clearpage

\begin{table}
\caption{ Lithium fractions: models and observations}
\begin{center}\scriptsize
\begin{tabular}{lcccccccc}\hline
Temperatures & $\tau_{min}$ & & Models & & & & Observations & \\
 & Gyrs.& $\alpha=0.0$ & $\alpha=0.5$ & $\alpha=1.0$ & $\alpha=1.5$ & $\alpha=2.0$ & Sp. Types & $F_{0.065}$ \\
\hline\hline 
      &  & Arizona & \\
$2050 \rightarrow 2700$ &$0.01$ & 14\% & 19\% &25\% &33\% & 45\% & M9.5$\rightarrow$M7   &  $6\pm4$\% \\
   & $0.05$ & 11 & 13& 16 & 20 & 25 \\
   & $0.1$ & 9 & 10.5& 11 & 15 & 17 \\
 & \\
$2050 \rightarrow 2500$ & $0.01$ & 18 & 23 & 30 & 40 & 51 & M9.5$\rightarrow$M8 &   $10\pm7$\% \\
   & $0.05$ & 15 &18 & 22 & 28 & 35 \\
   & $0.1$ & 14 &16 & 18 & 24 & 26 \\
\hline
      &  & Lyon & \\
$2050 \rightarrow 2700$ &$0.01$ & 8\% & 11\% &14.5\% &20\% & 28\% & M9.5$\rightarrow$M7   &  $6\pm4$\% \\
   & $0.05$ & 5.5 &7 & 9 & 11 & 14.5 \\
   & $0.1$ & 4 & 5& 6 & 7.5 & 9 \\
 & \\
$2050 \rightarrow 2500$ & $0.01$ & 10 & 14& 20 & 26 & 35 & M9.5$\rightarrow$M8 &    $10\pm7$\% \\
   & $0.05$ & 8 & 10 & 13 & 17 &  21\\
   & $0.1$ & 6.5 & 8 & 10 & 12.5 & 15 \\
\tableline      
\end{tabular}
\end{center}
Notes: predicted values of $F_{Li}$ for a K$_S<12$ sample of ultracool
dwarfs selected from a population with a power-law mass function, index $\alpha$,
and a constant birthrate over the period $\tau_{min} < \tau < 10$ Gyrs.

\end{table}
\clearpage

\begin{table}
\caption{Distances and kinematics}
\begin{center}\scriptsize
\begin{tabular}{lrcrrrcccc}\hline
2MASS & distance & Src. & M$_J$ & $\mu_\alpha$ & $\mu_\delta$ & Ref.& U & V & W \\
   & pc.  &  &  & $''$yr$^{-1}$ & $''$yr$^{-1}$ && km s$^{-1}$ & km s$^{-1}$& km s$^{-1}$ \\ 
\hline\hline
 2M0052-27 &  22.2$\pm   4.9$& 1&  11.96&  0.056& 0.090 & 1&   -10.4$\pm 1.1$&    4.6$\pm 0.7$&  -16.7$\pm 1.0$ \\ 
 2M0109+29 &  22.1$\pm   3.3$& 2&  11.20&  1.014 & 0.348 & 3 & -116.1$\pm 13.5$&  -18.1$\pm 8.5$&   18.0$\pm   0.7$ \\ 
 2M0140+27 &  17.9$\pm   2.7$& 2&  11.25&  0.061 & -0.252 &3&   -3.9$\pm 1.2$&  -10.2$\pm   0.1$&  -20.8$\pm 0.4$ \\ 
 2M0149+29 &  22.1$\pm   0.4$& 4&  11.69&  0.207 & -0.466 & 4&  -23.0$\pm 0.9$&  -17.6$\pm   0.4$&  -45.7$\pm 0.4$ \\ 
 2M0253+27 &  34.5$\pm   5.2$& 2&  10.55&  0.370 & 0.088 &3&  -68.6$\pm 6.1$&   -16.1$\pm 5.7$&   20.5$\pm  3.7$ \\ 
 2M0320+18 &  14.7$\pm   0.2$& 4&  10.90&  0.349 & -0.251 & 4&  -45.5$\pm 1.0$&  -19.5$\pm   0.0$&  -21.4$\pm 0.4$ \\ 
 2M0330+24 &  23.1$\pm   3.5$& 2&  10.54&  0.185&-0.039&3&  -43.1$\pm 2.3$&   -6.6$\pm 1.8$&   -8.2$\pm   1.3$ \\ 
 2M0335+23 &  23.5$\pm   3.5$& 2&  10.40&  0.058&-0.043&3&  -13.0$\pm 1.3$&   -4.8$\pm 0.4$&   -4.7$\pm   0.1$ \\ 
 2M0339-35 &   5.0$\pm   0.2$& 1&  12.25&  0.324&0.296&1&  -12.6$\pm 0.5$&   -6.2$\pm 0.8$&   -3.4$\pm 0.6$ \\ 
 2M0339+24 &  20.8$\pm   4.2$& 5&  11.27&  0.250&-0.010&6&  -40.8$\pm 3.1$&  -15.1$\pm 3.0$&   -5.8$\pm   2.6$ \\ 
 2M0350+18 &  24.9$\pm   3.7$& 2&  10.97&  0.189&-0.049&3&    4.8$\pm 2.3$&  -20.7$\pm 2.1$&   16.8$\pm   1.7$ \\ 
 2M0429+13 &  23.6$\pm   3.5$& 2&  10.81&  0.103&-0.016&3&  -43.5$\pm 1.4$&  -10.5$\pm 1.1$&   -9.7$\pm   0.9$ \\ 
 2M0746+20 &  12.1$\pm   0.3$& 4&  11.33& -0.358&-0.054&4&  -56.9$\pm 1.1$&  -15.9$\pm 0.2$&    0.4$\pm 0.1$ \\ 
 2M0810+14 &  13.8$\pm   2.1$& 2&  12.01&  -0.034&-0.128&3&  -20.5$\pm 1.0$&  -19.1$\pm 0.4$&    6.4$\pm   0.1$ \\ 
 2M0818+23 &  29.3$\pm   4.4$& 2&   9.99&  -0.275&-0.305&3&  -42.2$\pm 4.0$&  -44.7$\pm   0.6$&  -26.2$\pm 4.2$ \\ 
 2M0853-03 &   8.5$\pm   0.1$& 7&  10.43&  -0.509&-0.200&7&  -14.0$\pm 0.8$&  -10.8$\pm 0.7$&  -15.8$\pm   0.2$ \\ 
 2M0925+17 &  16.4$\pm   2.5$& 2&  11.53&  -0.232&0.010&3&  -20.8$\pm 2.6$&   -5.2$\pm 0.5$&   -4.0$\pm 1.2$ \\ 
 2M0952-19 &  24.5$\pm   3.7$& 2&  10.69&  -0.077&-0.104&3&    5.9$\pm 1.3$&   15.5$\pm 1.0$&  -24.9$\pm 0.4$ \\ 
 2M1016+27 &  19.3$\pm   2.9$& 2&  10.52&  -0.158&-0.461&3&   -8.8$\pm 2.3$&  -46.8$\pm 0.6$&    2.9$\pm 0.4$ \\ 
 2M1047+40A$^a$ &  24.7$\pm   3.7$& 2&   9.49&  -0.300&-0.030&8&  -31.4$\pm 5.0$&  -12.9$\pm 1.4$&  -10.8$\pm 1.5$ \\ 
 2M1047+40B$^a$ &  24.7$\pm   3.7$& 2&  10.39&  -0.300&-0.030&8&  -28.5$\pm 5.0$&  -13.0$\pm 1.4$&  -16.3$\pm 1.5$ \\ 
 2M1200+20 &  32.4$\pm   4.9$& 2&  10.30&  -0.150&0.232&3&  -34.8$\pm 3.3$&   26.3$\pm 1.8$&  -31.8$\pm   0.3$ \\ 
 2M1224-12 &  17.5$\pm   1.2$& 1&  11.34&  -0.262&-0.187&1&  -13.4$\pm 1.0$&  -17.3$\pm 1.4$&  -16.3$\pm   0.6$ \\ 
 2M1242+29$^b$ &  38.7$\pm   7.7$& 4&  11.49&  &    \\
 2M1246+31 &  22.6$\pm   3.4$& 2&  10.49&  -0.792&0.050&9&  -74.7$\pm 10.8$&  -41.1$\pm 6.8$&    5.6$\pm   0.9$ \\ 
 2M1253+40 &  20.6$\pm   3.1$& 2&  10.60&  0.225&-0.651&9&   52.3$\pm   2.6$&  -38.7$\pm   2.0$&   17.2$\pm   0.9$ \\ 
 2M1309-23 &  12.3$\pm   1.8$& 2&  11.32&  0.00& -0.210&6&   13.5$\pm   0.5$&  -18.7$\pm 0.6$&    2.9$\pm   0.6$ \\ 
 2M1336+47 &  22.5$\pm   3.4$& 2&  10.88&  0.111&-0.016&3&   19.5$\pm   2.7$&   12.2$\pm   2.6$&   -6.6$\pm   0.3$ \\ 
 2M1403+30 &  21.4$\pm   3.2$& 2&  11.04&  -0.788&0.042&3&  -67.6$\pm 8.4$&  -56.1$\pm 7.5$&  -15.1$\pm   4.4$ \\ 
 2M1501+22 &  10.6$\pm   0.8$& 3&  11.67&  -0.025&-0.058&4&    1.4$\pm   0.3$&   -1.5$\pm   0.1$&    5.6$\pm   1.0$ \\ 
 2M1504-23 &  13.3$\pm   2.7$& 4&  11.41&  -0.200&-0.060&6&  -35.3$\pm 0.6$&    0.4$\pm 2.0$&  -13.8$\pm   1.7$ \\ 
 2M1524+29 &  10.7$\pm   2.1$& 4&  11.07&  0.025&-0.480&6&   13.3$\pm   0.5$&  -20.8$\pm   0.6$&  -15.4$\pm   0.7$ \\ 
 2M1527+41 &  20.9$\pm   4.2$& 4&  10.88&  -0.170&-0.040&6&   -4.2$\pm 1.4$&  -11.7$\pm 1.7$&   12.4$\pm   2.7$ \\ 
 2M1550+30 &  24.2$\pm   4.8$& 4&  11.07&  -0.112&0.107&3&  -34.1$\pm 0.6$&  -23.0$\pm 1.2$&  -25.8$\pm   2.4$ \\ 
 2M1714+30 &  24.4$\pm   4.9$& 4&  11.00&  0.050&0.075&6&  -27.7$\pm   0.6$&  -20.4$\pm   1.3$&  -25.0$\pm 0.4$ \\ 
 2M2206-20 &  22.2$\pm   3.3$& 2&  10.70&  0.001&-0.065&3&   10.4$\pm   0.5$&   -0.8$\pm   0.3$&  -14.3$\pm 0.8$ \\ 
 2M2233+35 &  14.6$\pm   2.9$& 4&  11.12&  0.075&-0.250&6&    6.3$\pm 0.9$&  -21.4$\pm   0.7$&  -10.9$\pm 0.8$ \\ 
 2M2234+23 &  21.9$\pm   3.3$& 2&  11.44&  0.829&-0.034&3&  -69.4$\pm 10.8$&   -7.1$\pm 2.2$&  -53.3$\pm 6.8$ \\ 
 2M2235+18 &  25.5$\pm   3.8$& 2&  10.43&  0.326&0.042&3&  -37.1$\pm 4.9$&  -20.3$\pm 0.6$&   -6.9$\pm 3.4$ \\ 
\tableline      
\end{tabular}
\end{center}
Notes: \\
a. the two components of 2M1047+40 are treated as a single system in our analysis; \\
b. 2M1242+29 lacks proper motion measurements and is not included in the photometric sample. \\ 
Reference source for distance and astrometry: \\
1. trigonometric parallax and proper motions from Tinney (1996); \\
2. (M$_I$, (I-J) photometric parallax; 
3. proper motions from Gizis {\sl et al.} (2000a); \\
4. trigonometric parallax and proper motions from Dahn {\sl et al.} (2000);  \\
5. (M$_J$, (J-K$_S$) photometric parallax; \\
6. proper motions from POSS II/2MASS astrometry (this paper); \\
7. trigonometric parallax and proper motions from Monet {\sl et al.} (1992). \\
8. proper motions from Gizis {\sl et al.} (2000b). 
9. proper motions from  Luyten (1980)\\

\end{table}
\clearpage

\begin{figure}
\plotone{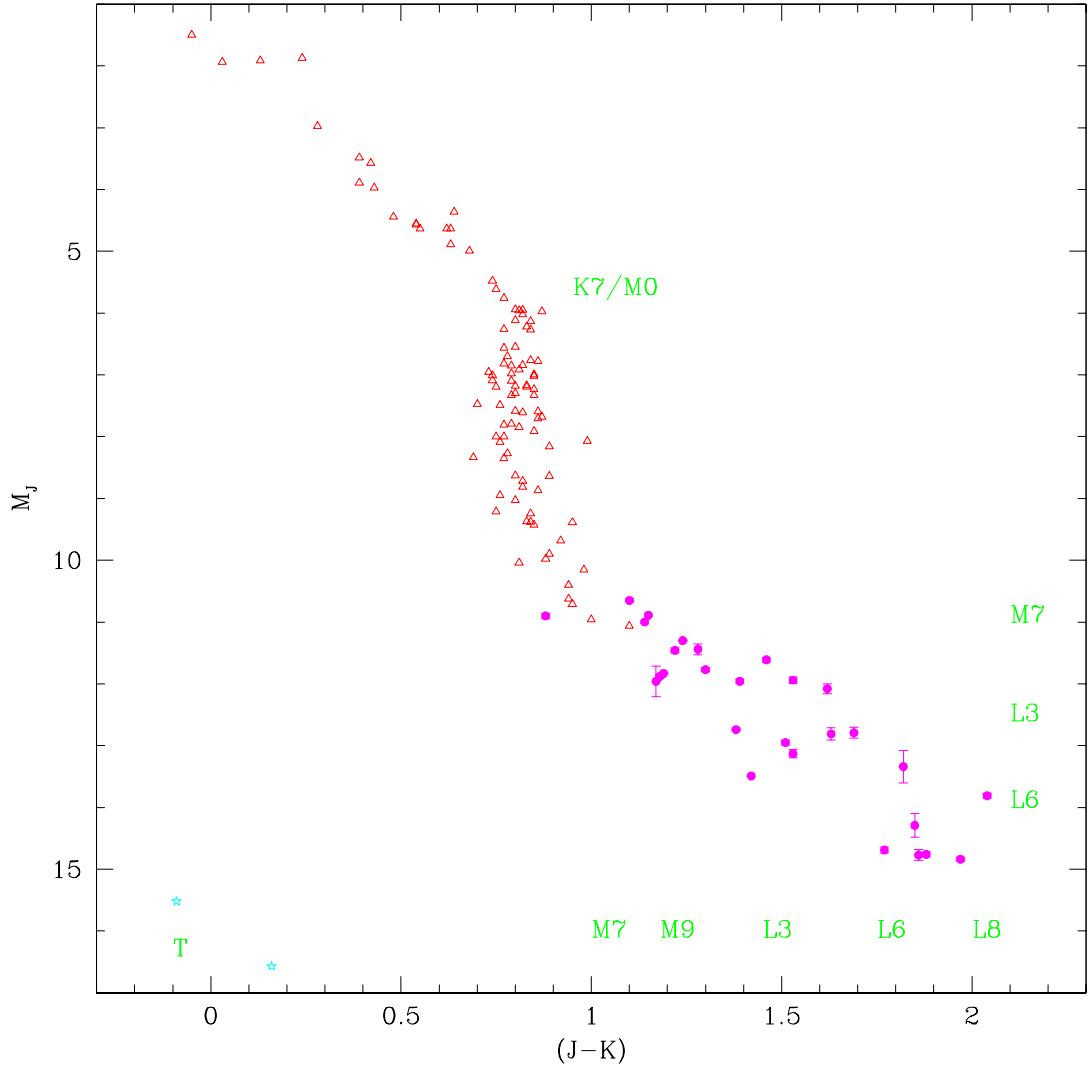}
\caption{ The (M$_J$, (J-K)) colour-magnitude diagram: crosses mark stars with 
trigonometric parallax measurements by
the Hipparcos satellite (ESA, 1997), while open triangle plot data for stars within 
8 parsecs of the Sun
(R99). Photometry for those stars is taken from Leggett (1992) and the 2MASS
 database. The solid points identify ultracool dwarfs with JHK photometry 
and parallax measurements by Dahn {\sl et al.} (2000). }
\end{figure}

\clearpage
\begin{figure}
\plotone{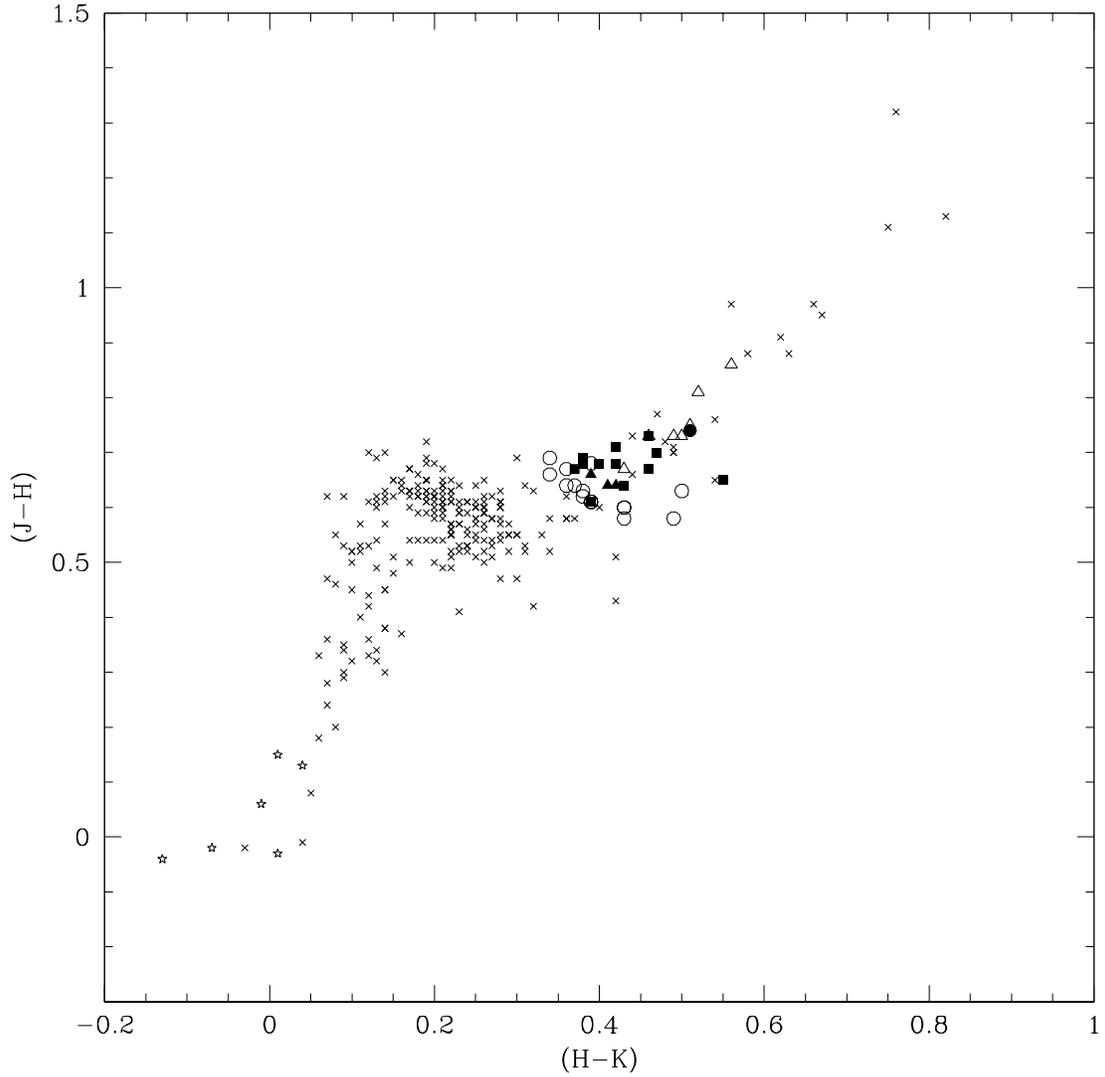}
\caption {The ((J-H), (H-K$_S$) diagram: crosses mark data for stars and L dwarfs with known
parallax; T dwarfs are plotted as 5-points stars; ultracool dwarfs from Table 1 with
spectral types M6/M6.5 are plotted as solid triangles, M7/M7.5 as open circles, M8/M8.5 as solid
squares, M9/M9.5 as open triangles, and 2M0746+20 (L0.5) as a solid point. The (J-K$_S$)
selection criterion is obvious from the lower boundary of the sample.}
\end{figure}

\clearpage
\begin{figure}
\plotone{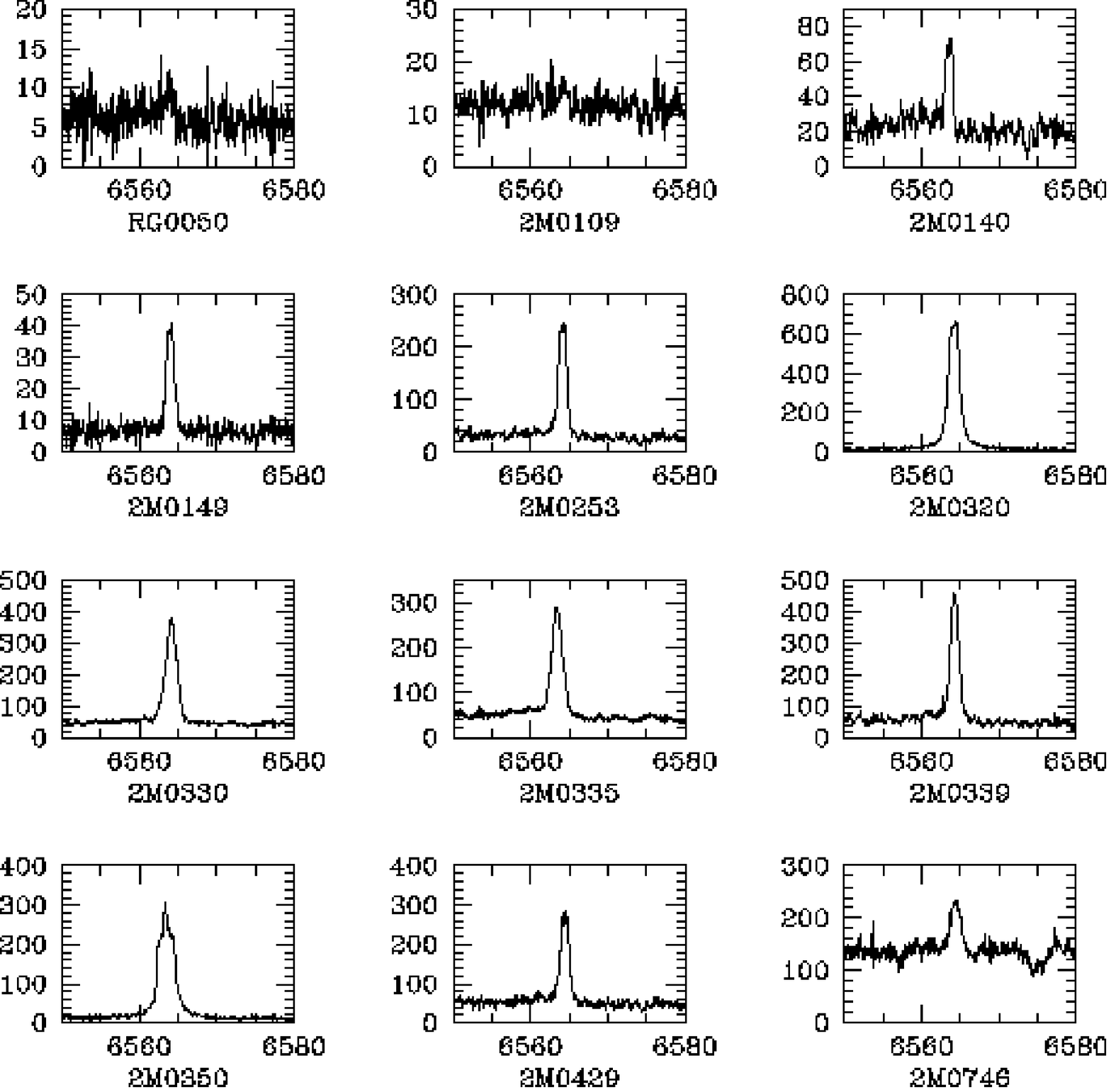}
\caption {High-resolution observations of the H$\alpha$ region for ultracool dwarfs in the
present sample.}
\end{figure}

\clearpage
\begin{figure}
\plotone{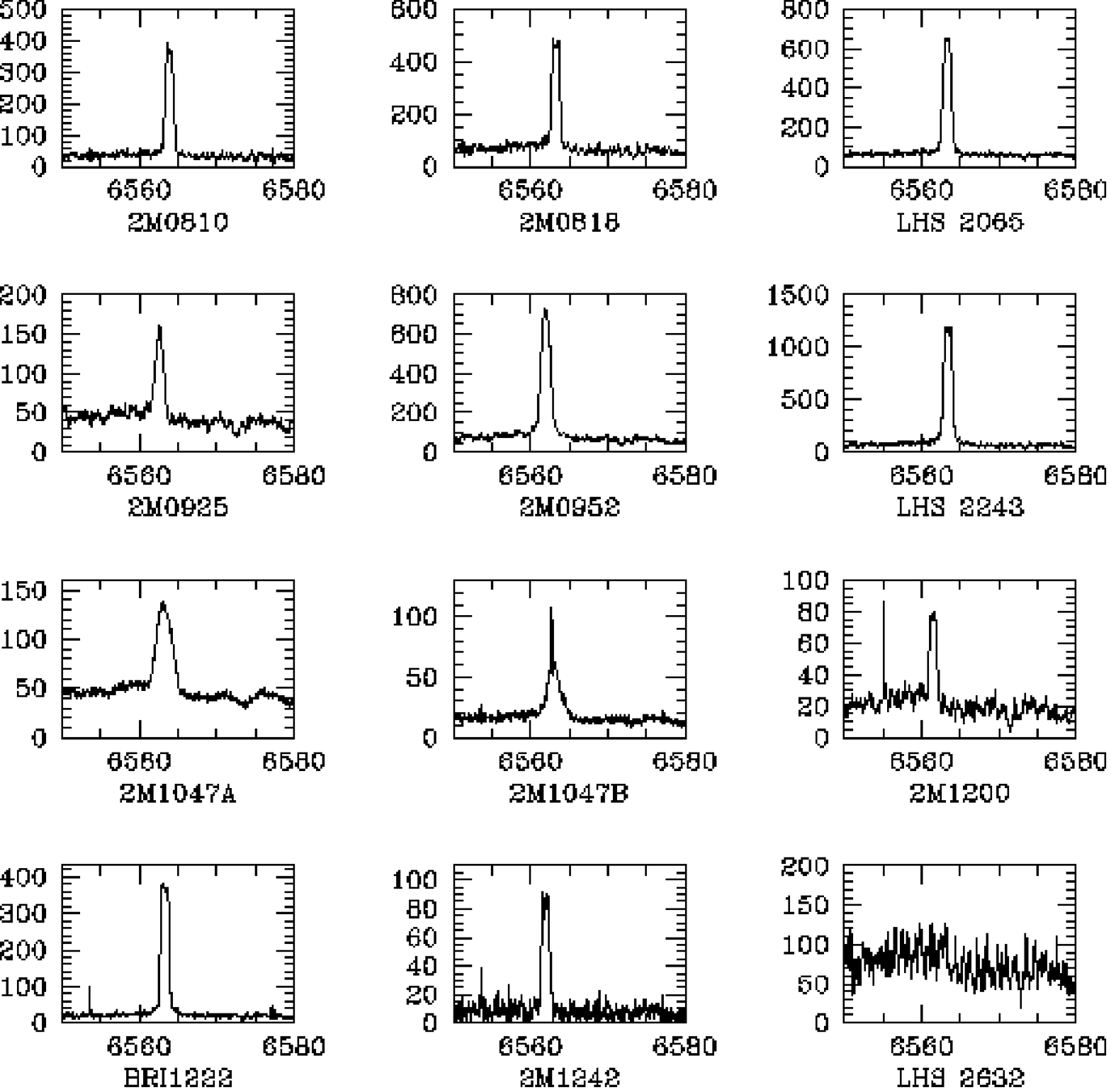}
\caption {High-resolution observations of the H$\alpha$ region for ultracool dwarfs in the
present sample (contd.).}
\end{figure}

\clearpage
\begin{figure}
\plotone{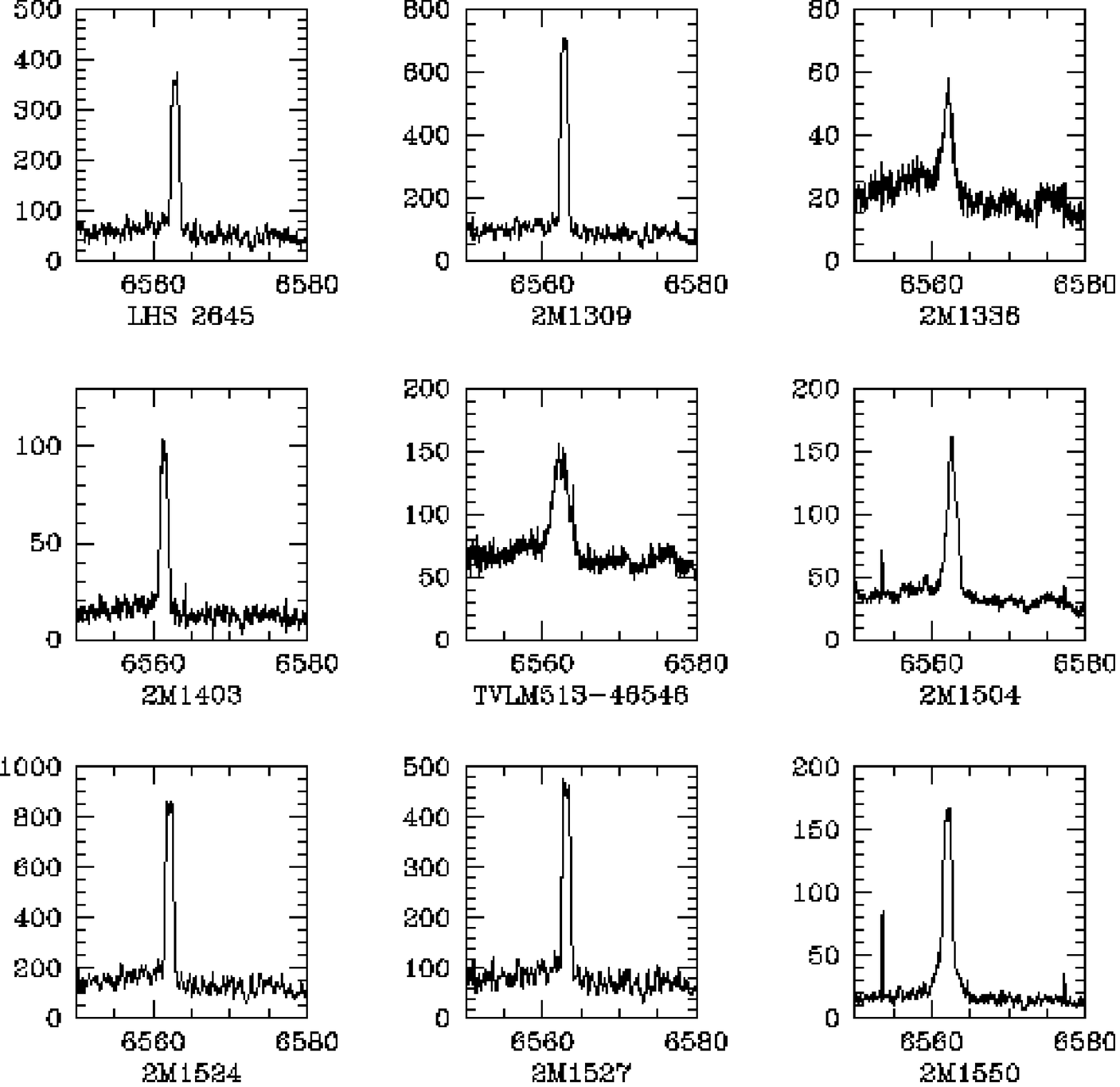}
\caption {High-resolution observations of the H$\alpha$ region for ultracool dwarfs in the
present sample (contd.).}
\end{figure}
\clearpage

\begin{figure}
\plotone{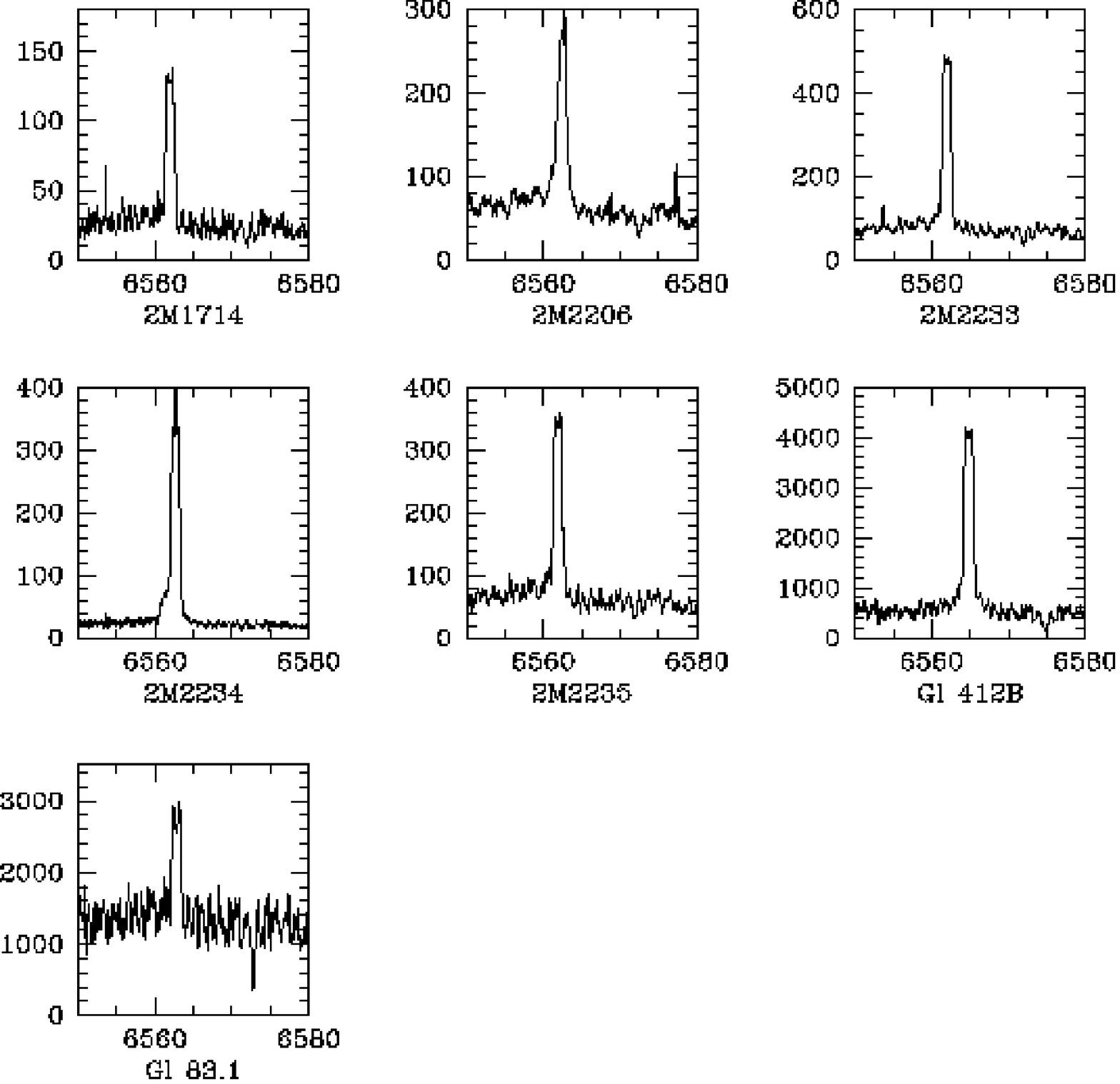}
\caption {High-resolution observations of the H$\alpha$ region for ultracool dwarfs in the
present sample (contd.).}
\end{figure}

\clearpage
\begin{figure}
\plotone{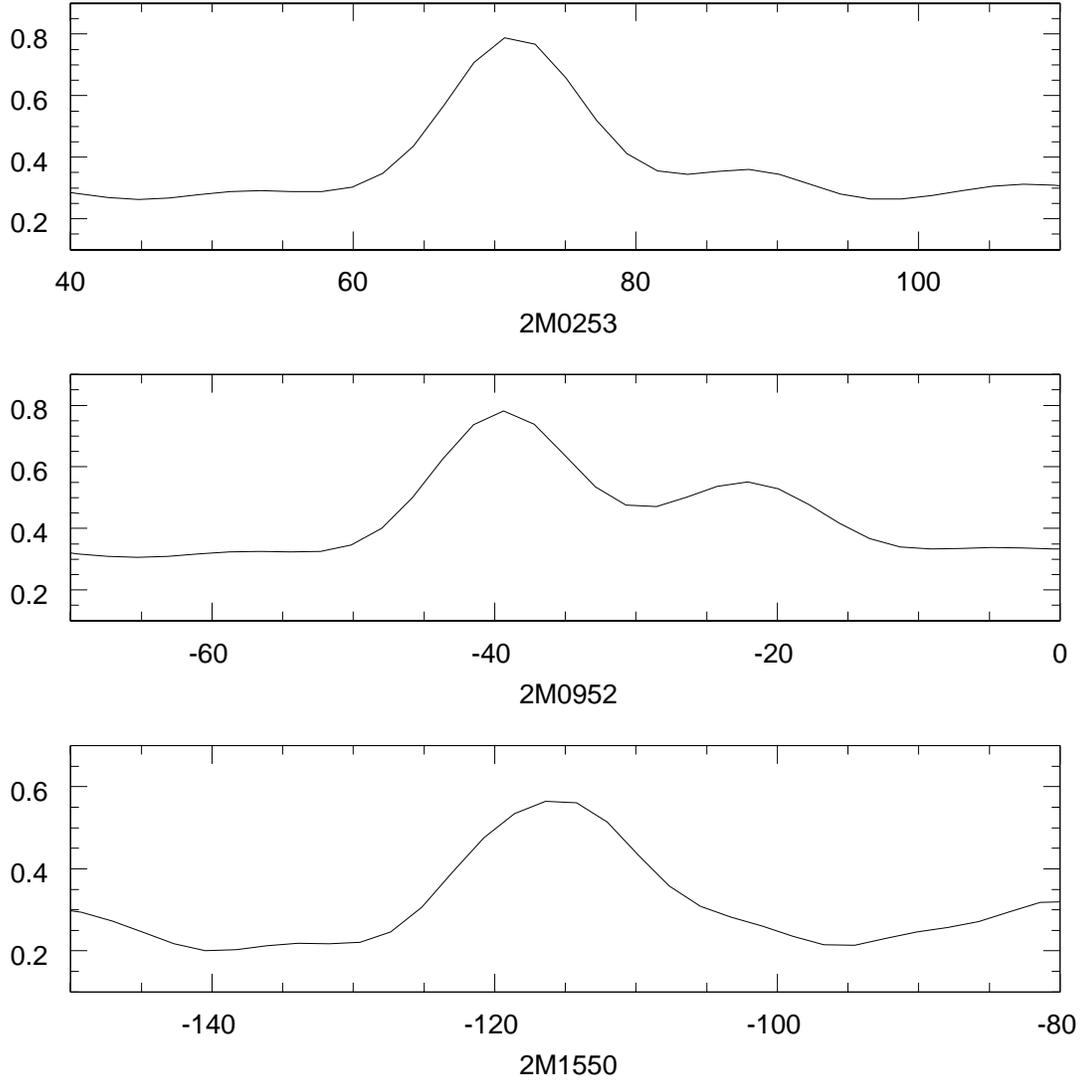}
\caption {Representative cross-correlation spectra for 2M0253+27, 2M0952+17 and 2M1550+30.
The first two dwarfs show bimodal distributions, characteristics of
double-lined spectroscopic binaries, while the last has an asymmetric profile,
suggestive of a high flux-ratio SB2.}  
\end{figure}

\begin{figure}
\plotone{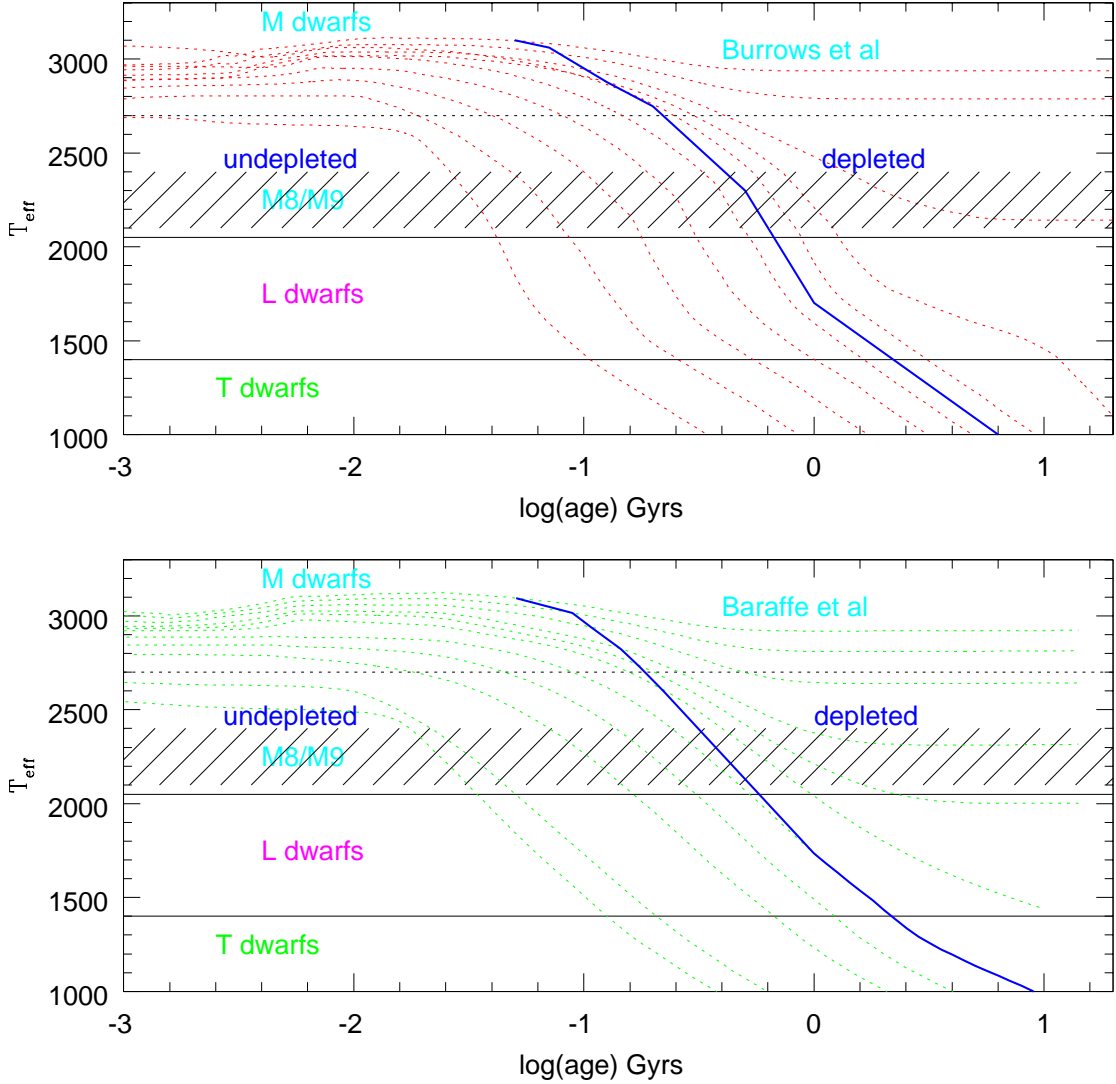}
\caption{ Lithium depletion for low-mass stars and brown dwarfs. The upper panel
plots evolutionary tracks from the Burrows {\sl  et al.} (1997) models for masses
of 0.02, 0.03, 0.04, 0.05, 0.06, 0.07, 0.075, 0.08, 0.09 and 0.1 M$_\odot$; the
lower panel plots tracks for masses of 0.02, 0.025, 0.04, 0.05, 0.06, 0.07, 0.075, 0.08, 0.09,
0.10 and 0.11 M$_\odot$ from the Baraffe {\sl  et al.} (1998) calculations. Note that the
stellar/brown dwarf transition occurs at $\sim0.075M_\odot$ in the former models,  and at
$\sim0.072 M_\odot$ in the latter.
In both cases, we delineate the approximate boundaries of the M dwarf, L dwarf and T dwarf r\'egimes;
the shaded region marks the appropriate temperature limits for spectral types M8 to M9.5
(see text for discussion). The heavy solid lines outline the R$_{Li} = 1\%$ boundary;
lithium should not be detected in dwarfs which have evolved beyond this limit.}
\end{figure}

\begin{figure}
\plotone{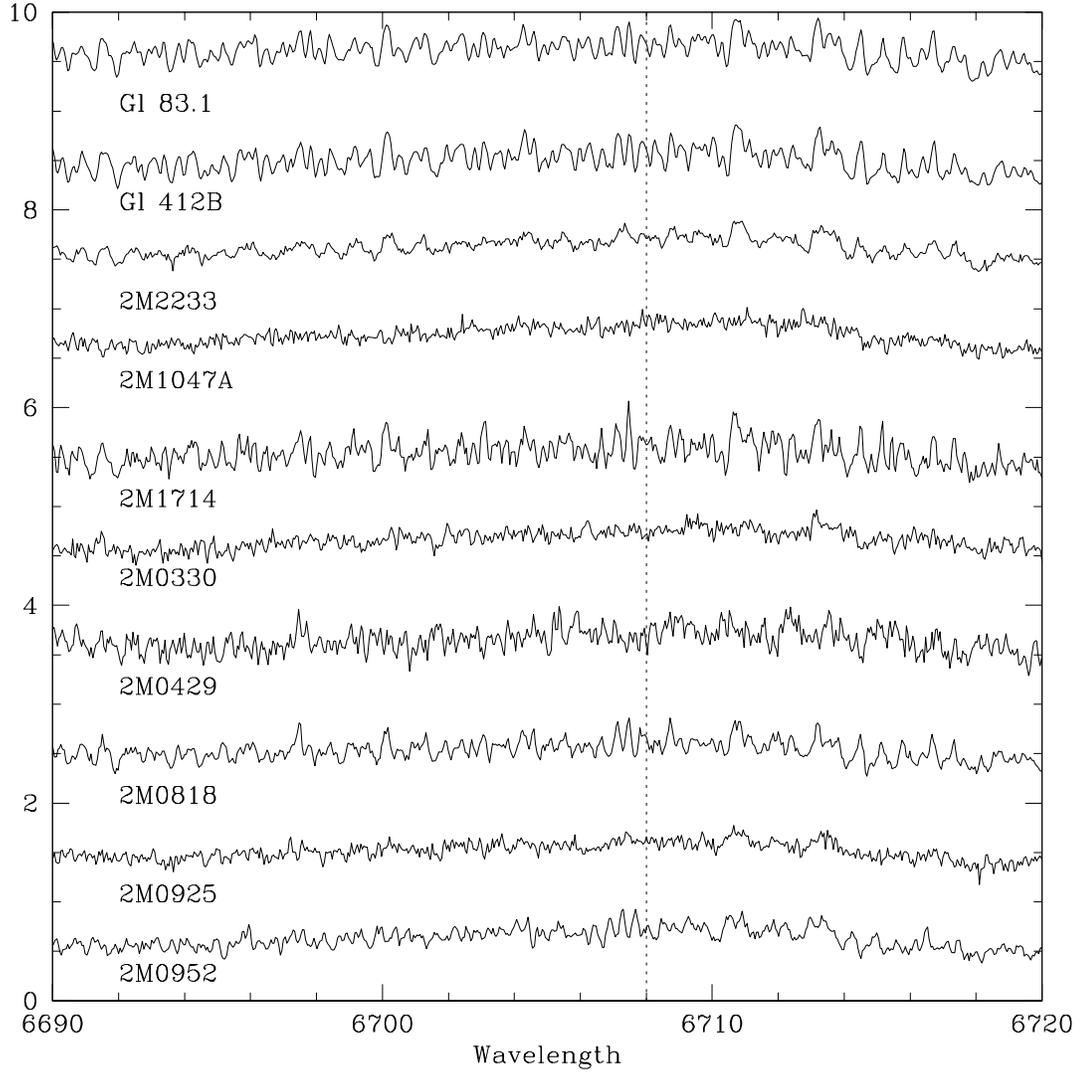}
\caption{ Lithium in ultracool dwarfs: all of the spectra have been
adjust to align any potential Li I absorption with the dotted line. This
figure plots data for (from top to bottom) Gl 83.1 (M4.5); Gl 412B (M5);
2M2233+35 (M6); 
2M1047+40A and 2M1714+30 (M6.5); 2M0330+24, 2M0429+13, 2M0818+23, 2M0925+17 
and 2M0952-19 (M7).}
\end{figure}

\begin{figure}
\plotone{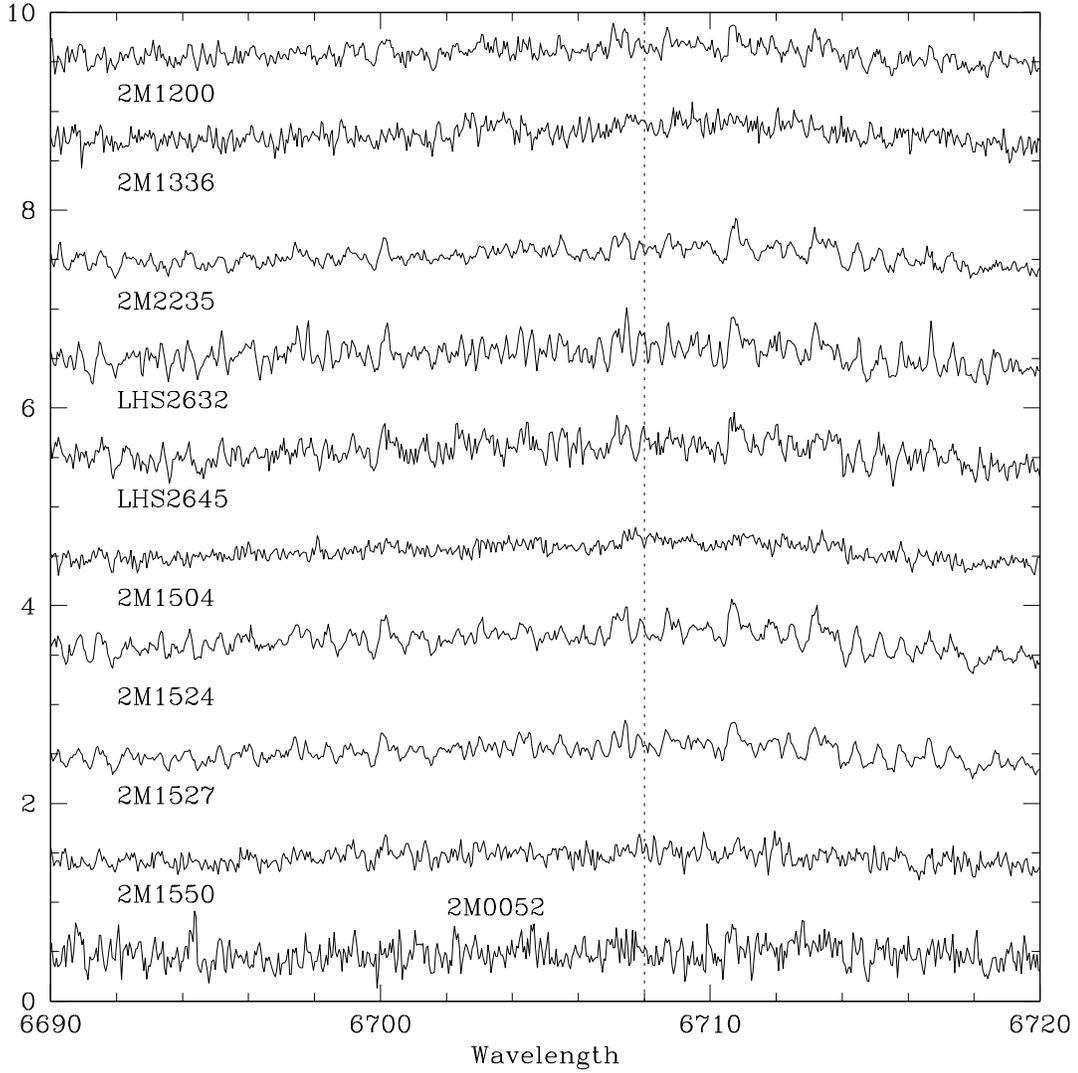}
\caption{ Lithium in ultracool dwarfs (contd.): M7 and M8 dwarfs, including
(from top to bottom)  2M1200+20, 2M1336+47, 
2M2235+18 (M7); LHS 2632, LHS 2645, 2M1504-23, 2M1524+29,
2M1527+41 and 2M1550+30 (M7.5); and 2M0052-27 (M8)} 
\end{figure}

\begin{figure}
\plotone{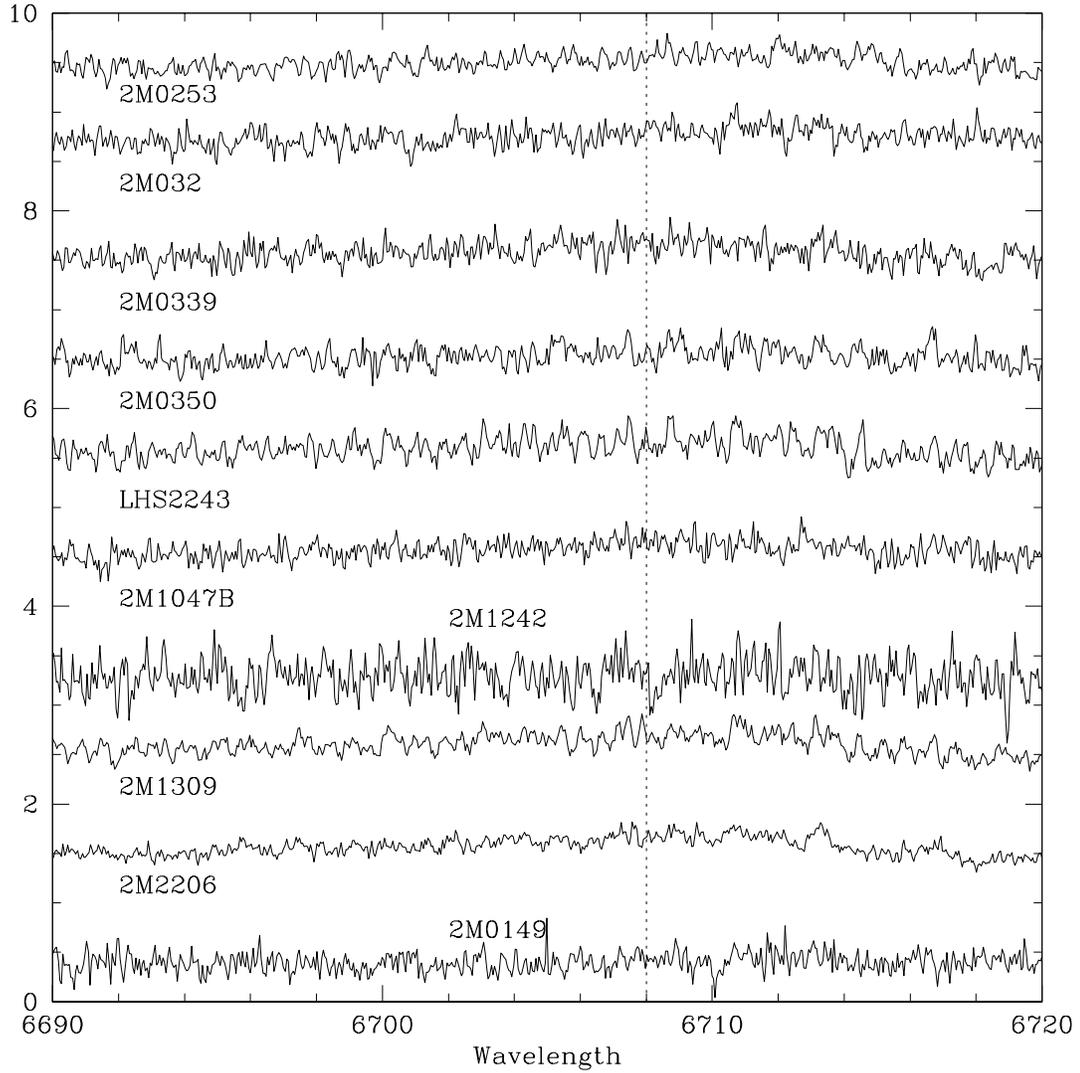}
\caption{ Lithium in ultracool dwarfs (contd.): M8 and M8.5 dwarfs, including
(from top to bottom) 2M0253+27, 2M0320+18, 2M0339+24, 2M0350+18, 
LHS 2243, 2M1047+40B, 2M1242+29, 2M1309-23 and 2M2206-20 (M8); and 
2M0149+29 (M8.5). 2M1242+29 may exhibit lithium absorption, although the signal-to-noise
is low. }
\end{figure}

\begin{figure}
\plotone{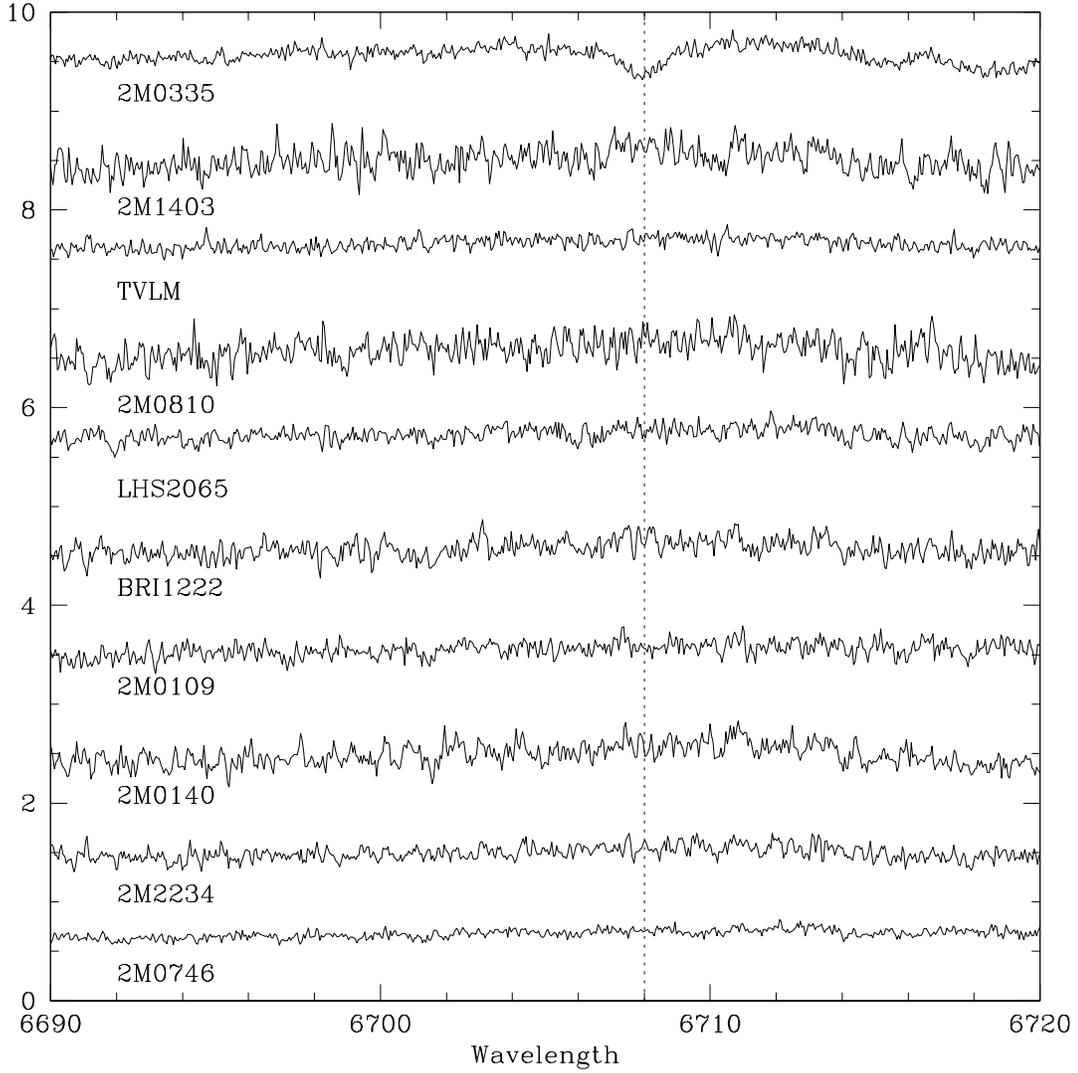}
\caption{ Lithium in ultracool dwarfs (contd.): the latest-type dwarfs in the present sample,
including 2M0335+23, 2M1403+30 and TVLM513-46546 (M8.5); 2M0810+14, LHS 2065 and 2M1224-12 (M9); 
2M0109+29, 2M0140+27 and 2M2234+35 (M9.5); and 2M0746+20 (L0.5).
Lithium is clearly present in 2M0335+23.}
\end{figure}

\begin{figure}
\plotone{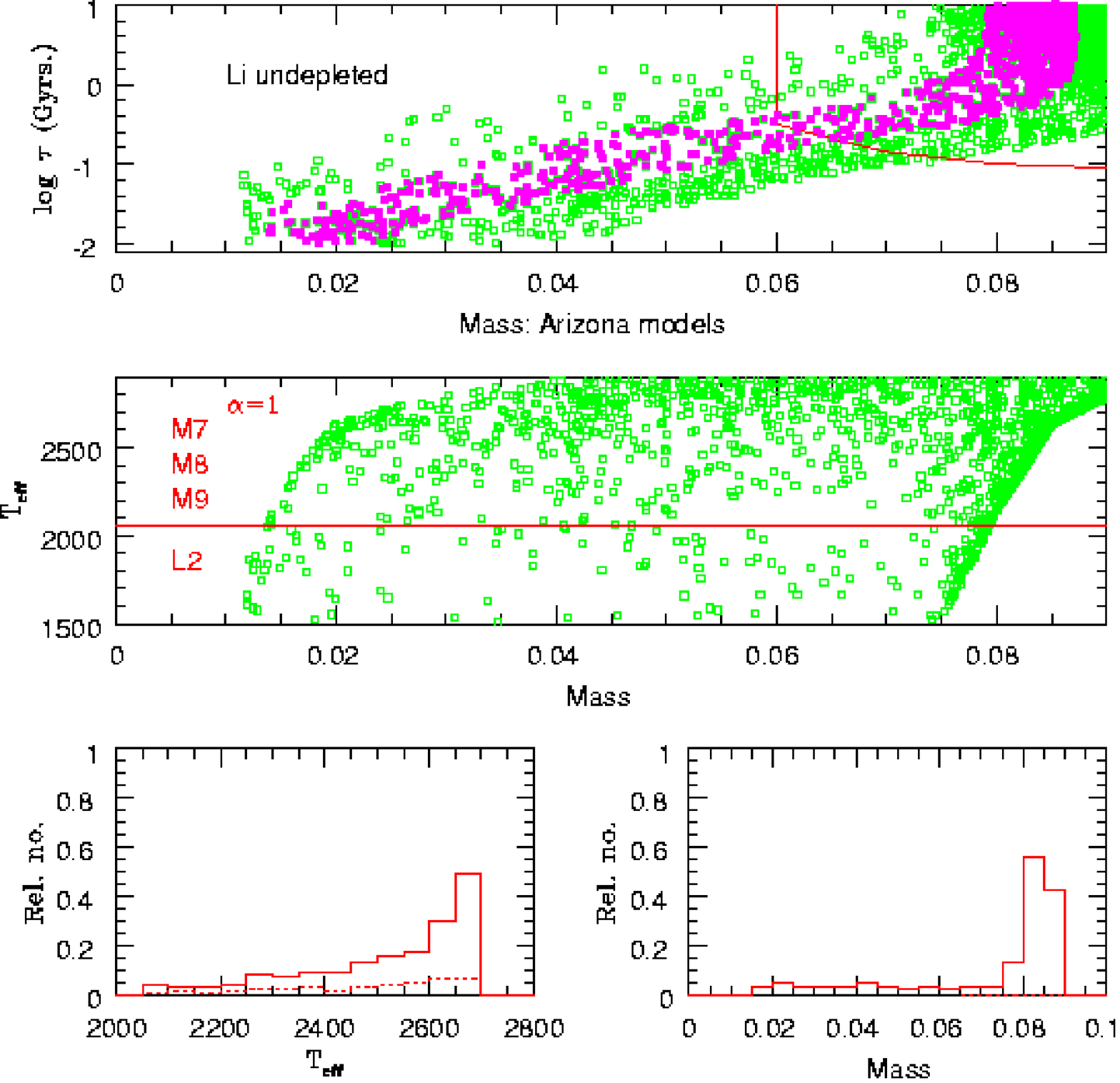}
\caption{ The (mass, T$_{eff}$) and (mass, age) distributions predicted 
for low-mass dwarfs with K$_S \le 12$, drawn for a mass function 
$\Psi(M) \propto M^{-1}$, based on the Arizona theoretical tracks. 
In the uppermost diagram, the filled squares identify
dwarfs in the temperature range $2700 > T_{eff} > 2050$K; the solid line marks the
approximate location of the lithium depletion boundary. The histograms plot the
number distribution for the full K$_S<12$ simulation as a function of temperature
and mass; the dotted histogram marks the contribution from lithium-rich dwarfs, and
the histograms are scaled to match the corresponding distributions in Figure 14. }
\end{figure}

\begin{figure}
\plotone{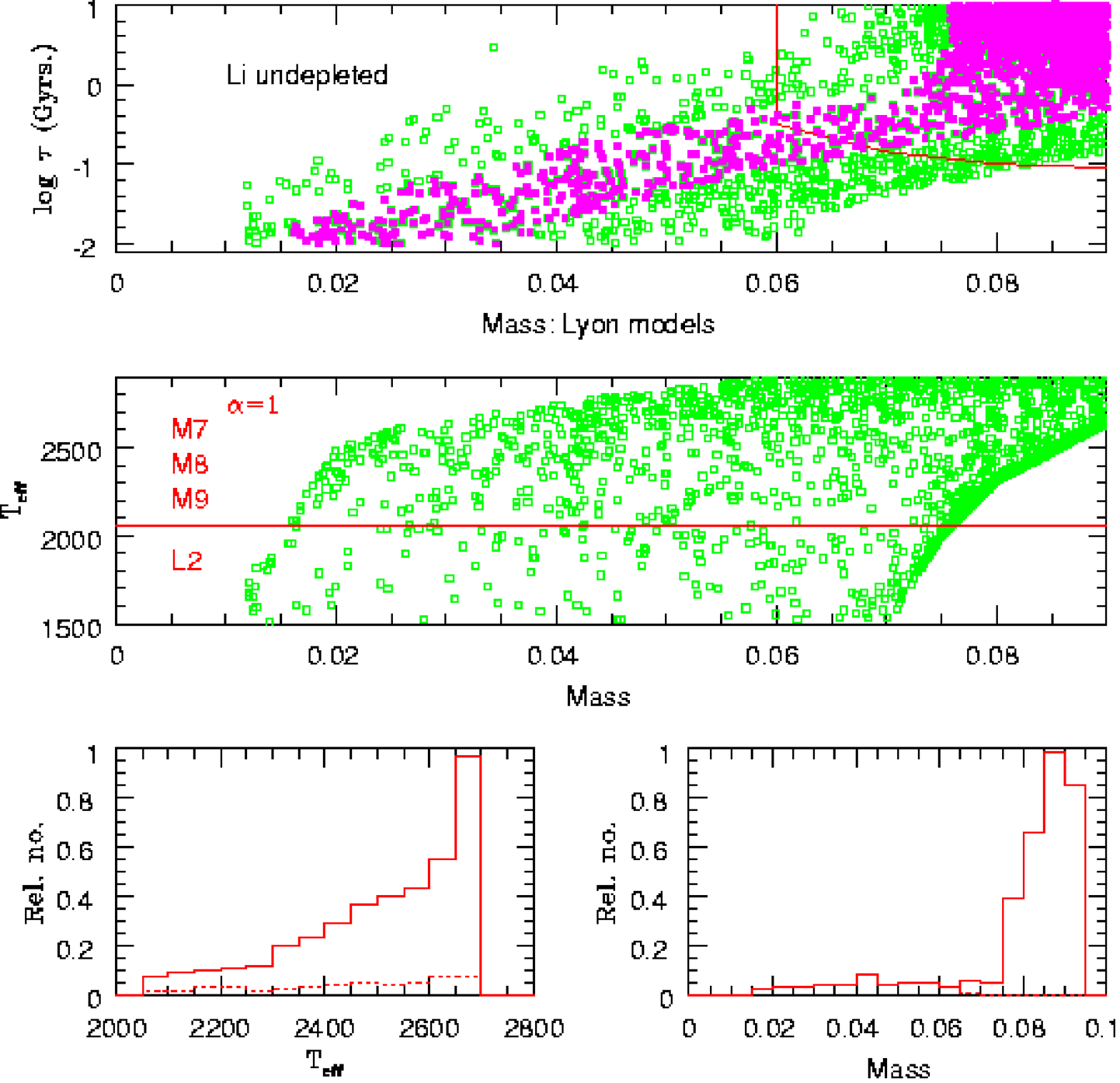}
\caption{  The (mass, T$_{eff}$) and (mass, age) distributions  predicted 
for low-mass dwarfs with K$_S \le 12$, drawn for a mass function 
$\Psi(M) \propto M^{-1}$, based on the Lyon theoretical tracks. As in Figure 13,
 filled squares in the uppermost diagram identify
dwarfs in the temperature range $2700 > T_{eff} > 2050$K, and the solid line marks the
approximate location of the lithium depletion boundary. The scaling of the two histograms,
plotting the temperature and mass distribution of the full sample, matches that in Figure 13.
Note the larger number of stellar-mass dwarfs predicted by the Lyon models.  }
\end{figure}

\begin{figure}
\plotone{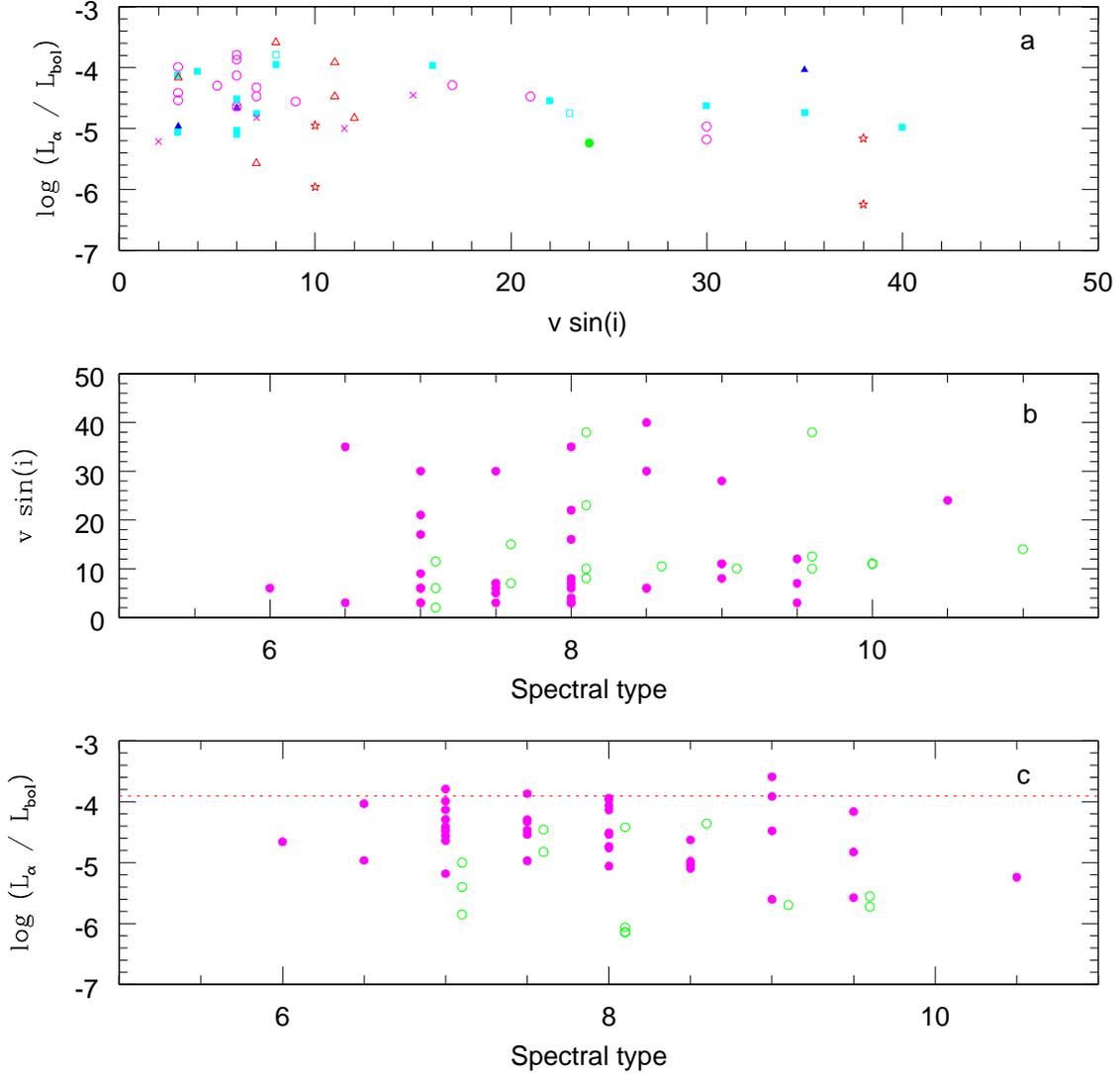}
\caption{Activity and rotation in ultracool dwarfs. Panel $a$ plots activity, 
log$_{10}(L_\alpha/L_{bol})$,
as a function of measured rotation, $v \sin{i}$, for both our observations of ultracool dwarfs and
M7-M9.5 dwarfs from Basri (2001). In the case of stars in common between the samples, we use our data.  
We use different symbols for each spectral type: for our data, the
symbols have the same meaning as in figure 2; M7 dwarfs from Basri's sample are plotted as crosses, 
M8/M8.5 dwarfs as open squares and M9/M9.5 dwarfs as  5-point stars. Panel $b$ plots rotation as
a function of spectral type, and panel $c$ plots activity as a function of spectral type, 
where n=5 is spectral type M5, n=10 is spectral type L0. The dotted horizontal in the
latter panel marks the mean activity level amongst earlier-type M dwarfs in the field.  
In both cases, the solid points are from this paper and the open circles from Basri's
observations, and, for clarity, we offset the spectral types by +0.1 for the latter sample.}
\end{figure}

\begin{figure}
\plotone{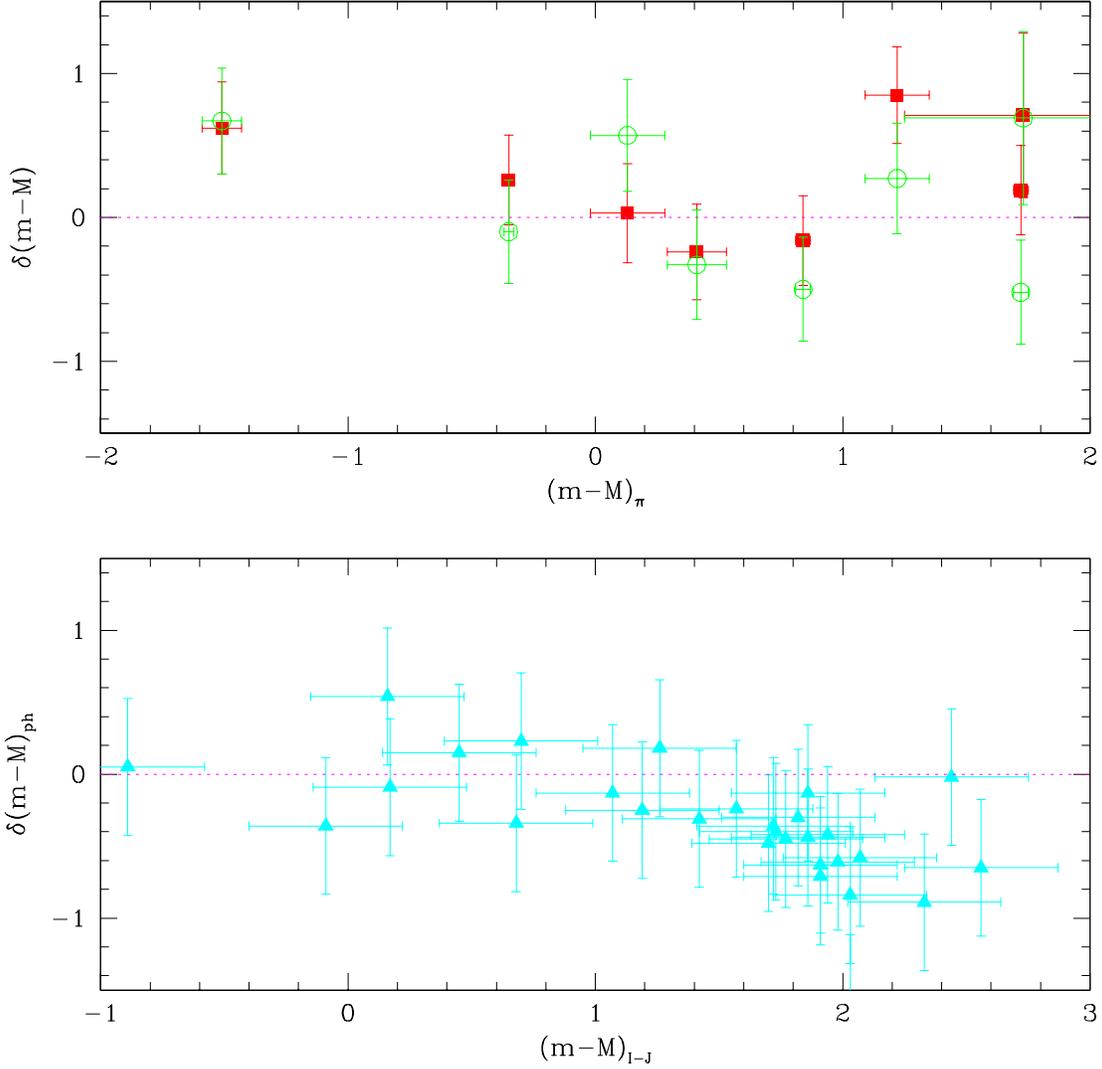}
\caption{Comparison of parallax determinations for ultracool dwarfs. The upper
panel compares trigonometric and photometric distance modulus estimates for dwarfs in the current sample
with astrometric measurements, where $\delta(m-M) = (m-M)_{phot} - (m-M)_{trig}$; solid squares
plot the comparison with (m-M)$_{I-J}$, open circles show (m-M)$_{J-K}$. The lower panel
compares distance moduli derived from the photometric parallaxes, where  
$\delta(m-M)_{ph} = (m-M)_{J-K} - (m-M)_{I-J}$. }
\end{figure}

\begin{figure}
\plotone{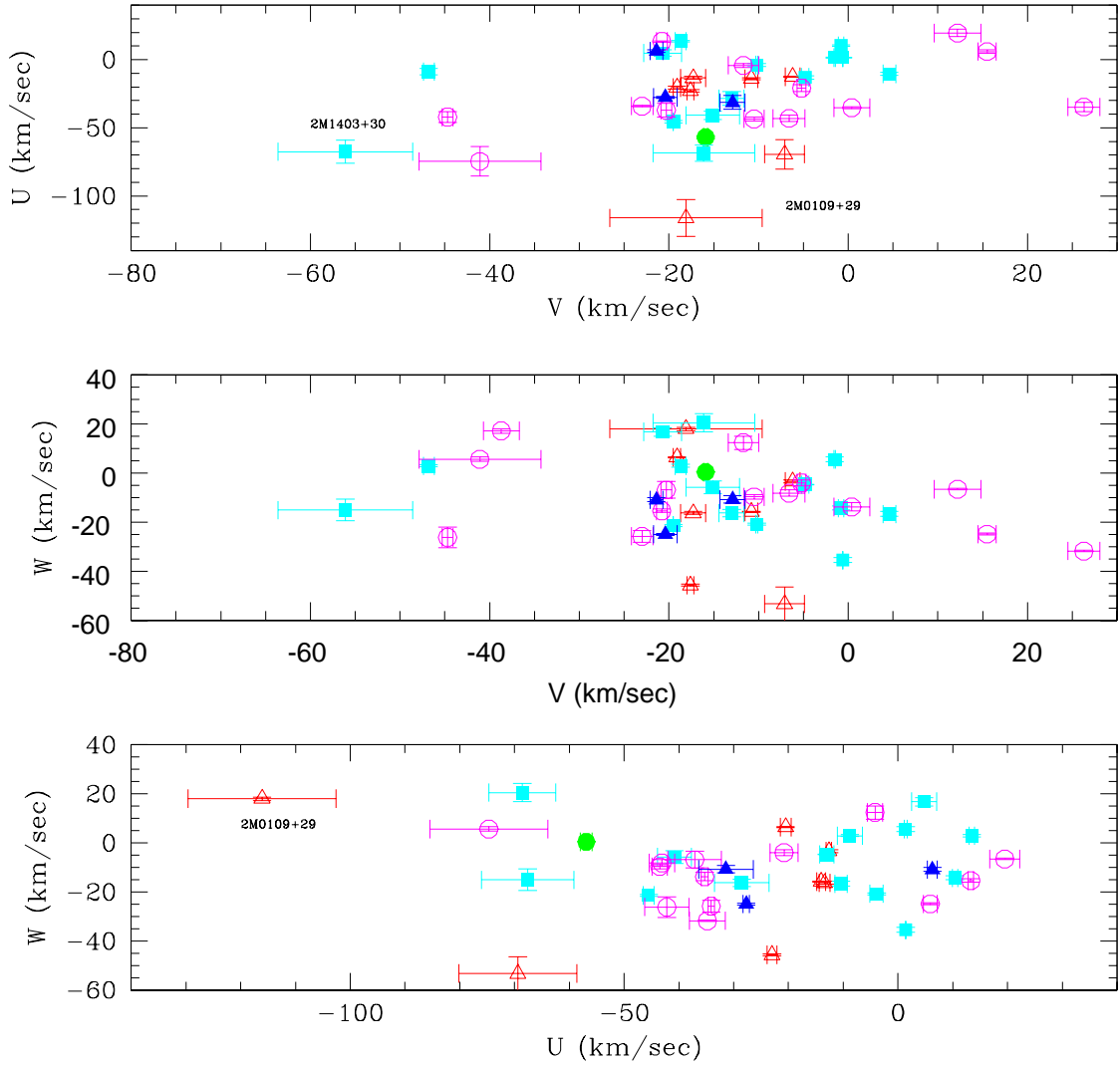}
\caption{Space motions of ultracool dwarfs: the symbols are coded by spectral type as in figures
2 and 10, and the two most extreme outliers identified.}
\end{figure}

\begin{figure}
\plotone{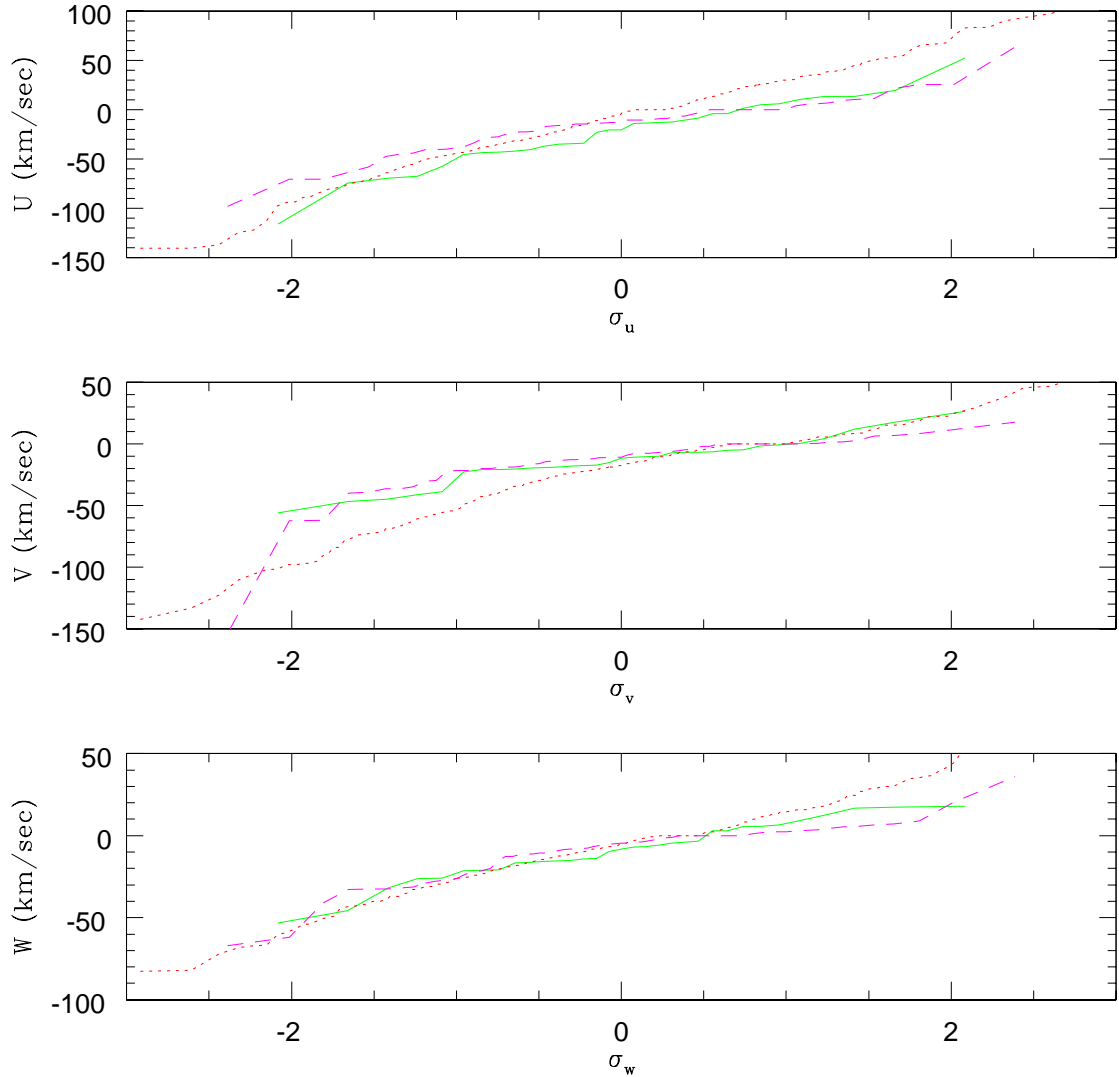}
\caption{Probability plots, comparing the velocity distribution of the the ultracool dwarfs
in the present sample against data for dM and dMe stars from the PMSU survey (Reid {\sl et al.}, 1995). 
The solid line plots data for the
photometrically-selected ultracool dwarf sample; the dotted line shows the distribution of
dM dwarfs; and the dashed line plots data for the dMe sample. }
\end{figure}


\begin{thebibliography}{}

\bibitem[Barrado y Navascu{\' e}s, Stauffer, Bouvier, \& Mart{\' i}n(2001)]{2001ApJ...546.1006B} Barrado y Navascu{\' e}s, D., Stauffer, J.~R., Bouvier, J.~;., \& \pedant, E.~L.\ 2001, \apj, 546, 1006
\bibitem[Baraffe et al, 1998] {bcah98} Baraffe, I., Chabrier, G., Allard, F., Hauschildt, P.H. 1998, 
\aap, 337, 403
\bibitem[Basri, 2001] {nas1} Basri, G. 2001, Cool Stars 11 Workshop, ed. Ramon J. Garcia Lopez,
Rafael Rebolo \& Maria Rosa Zapatero Osorio, ASP Conf. Ser. 223, 261
\bibitem[Basri et al.(2000)]{2000ApJ...538..363B} Basri, G., Mohanty, S., Allard, F., Hauschildt, 
P.~H., Delfosse, X., \pedant, E.~L., Forveille, T., \& Goldman, B.\ 2000, \apj, 538, 363
\bibitem[Becklin et al, 1969] {be69} Becklin, E.E., Frogel, J.A., Hyland, A.R.,
Kristian, J., Neugebauer, G. 1969, \apj, 158, L133
\bibitem[Bessell, 1991]{b91} Bessell, M.S. 1991, \aj, 101, 662
\bibitem[Burrows \& Liebert, 1993] {bl93} Burrows, A., Liebert, J. 1993, Rev. Mod. Phys. 65, 301
\bibitem[Burrows et al, 1997] {bur97} Burrows, A., Marley, M., Hubbard, W.B., Lunine, J.I., 
Guillot, T., Saumon, D., Freedman, R., Sudarsky, D., Sharp, C. 1997, \apj, 491, 856
\bibitem[Burgasser et al, 1999] {bg99}  Burgasser, A., Kirkpatrick, J.D., Brown, M.E., Reid, I.N., 
Gizis, J.E. {\sl et al.} 1999,  \apj, 522, L65
\bibitem[Chabrier \& baraffe, 1997] {cb97} Chabrier, G., Baraffe, I. 1997, \aap, 327, 1039
\bibitem[Dahn {\sl et al.}2000] {d20} Dahn, C.C., Guetter, H.H., Harris, H.C., {\sl  et al.} 2000, 
in {\sl From Giant Planets to Cool Stars}, ed. C.A. Griffiths \& M.S. Marley, ASP Conf. Ser. Vol. 212, 74 
\bibitem [Dahn {\sl et al.}, 2002] {da20} Dahn, C.C., Harris, H.C., Vrba, F.J., Guetter, H.H.,
{\sl et al.} 2002, \aj, submitted.
\bibitem[Delfosse et al, 1998] {del98} Delfosse, X., Forveille, T., Perrier, C.,
Mayor, M. 1998, A\& A, 331, 581
\bibitem[Durney et al, 1993] {du93} Durney, B.R., DeYoung, D.S., Roxburgh, I.W. 1993, 
{\sl Solar Phys.}, 145, 207
\bibitem[Epchtein {\sl et al.}, 1994] {ep94} Epchtein, N., De Batz, B., Copet, E. {\sl et al.} 1994,
Ap.Sp.Sci., 217, 3
\bibitem[ESA, 1997] {esa97} ESA, 1997, The Hipparcos Catalogue, ESA SP-1200
\bibitem [Gizis et al, 2000] {g20}   Gizis, J.E., Monet, D.G., Reid, I.N., Kirkpatrick, J.D., 
Liebert, J., Williams, R.J. 2000a AJ, 120, 1085 (G2000)
\bibitem [Gizis et al, 2000] {g20b} Gizis, J.E., Monet, D.G.,  Reid, I.N., Kirkpatrick, J.D., 
Burgasser, A.B. 2000b, MNRAS, 311, 385
\bibitem [Gizis et al, 2002] {giz2002} Gizis, J.E., Reid, I.N., Hawley, S.L. 2002, \aj, in press (PMSU3)
\bibitem [Gliese, 1969] {cns} Gliese, W. 1969,  Veroff. Astr. Rechen-Instituts, Heidelberg, Nr. 22
\bibitem [Gray, 1982] {gr82} Gray, D.F. 1982,  {\sl The observation and analysis of stellar 
photospheres}, Cambridge University Press, (New York, Cambridge)  
\bibitem[Hawley et al, 1996] {haw96} Hawley, S.L., Gizis, J.E., Reid, I.N. 1996, \aj, 112, 2799 [PMSU2]
\bibitem[Hawley et al] {haw} Hawley, S.L., Reid, I.N. ,Gizis, J. 2000, in {From Giant Planets to
Cool Stars}, ed. C.A. Griffith \& M.S. Marley, ASP Conf. Ser. 212, p. 252
\bibitem[Hawkins \& Bessell, 1988] {hb88} Hawkins, M.R.S., Bessell. M.S 1988, MNRAS, 234, 177
\bibitem[Henry et al, 1995] {he95} Henry, T.J., Kirkpatrick, J.D., Simons, D.A. 1995, AJ, 108, 1437
\bibitem[Henry et al, 1996] {he96} Henry, T.J., Soderblom, D.R., Donahue, R.A., Baliunas, S.L. 1996,
\aj, 111, 439
\bibitem[Henry et al, 1997] {he97} Henry, T.J., Ianna, P.A., Kirkpatrick, J.D., Jahrei{\ss}, H. 1997,
\aj, 114, 388
\bibitem[Herbig, 1956]{h56} Herbig, G. 1956, \pasp, 68, 531
\bibitem[Hyland et al, 1969] {hy69} Hyland, A.R., Becklin, E.E., Neugebauer, G., 
Wallerstein, G. 1969, \apj, 158, 619
\bibitem [Jahreiss \& Wielen]{jw} Jahrei{\ss}, H., Wielen, R. 1983, in IAU Colloquium 76,
{\sl The Nearby Stars and the Luminosity Function}, ed. A.G. Davis Philip \& A.R.
Upgren, L. Davis Press, Schenectady, New York, p. 277
\bibitem[Kastner et al, 1997] {ka97} Kastner, J.H., Zuckerman, B., Weintraub, D.A., Forveille, T.
1997, Science, 277, 67
\bibitem[Kenworthy et al, 2001] {ken20} \bibitem[Kenworthy et al.(2001)]{2001ApJ...554L..67K} Kenworthy, M.~et al.\ 
2001, \apjl, 554, L67. 
\bibitem[Kirkpatrick et al, 1994] {k94} Kirkpatrick, J.D., McGraw, J.T., Hess, T.R., Liebert, J., McCarthy, D.W.
1994, \apjs, 94, 749
\bibitem[Kirkpatrick et al, 1995] {k95} Kirkpatrick, J.D., Henry, T.J., Simons, D.A. 1995, \aj, 109, 797
\bibitem[Kirkpatrick et al, 1997] {k97} Kirkpatrick, J.D., Henry, T.J., Irwin, M.J. 1997, \aj, 113, 1421
\bibitem[Kirkpatrick et al, 1999] {k99} Kirkpatrick, J.D., Reid, I.N., Liebert, J., Cutri, R. {\sl et al.}
1999, ApJ, 519, 802
\bibitem[Kirkpatrick et al, 2000] {k20}  Kirkpatrick, J.D., Reid, I.N., Liebert, J., Gizis, J.E. {\sl  et al.}
2000, AJ, 120, 447
\bibitem[Lane et al.(2001)]{2001ApJ...560..390L} Lane, B.~F., Zapatero Osorio, M.~R., Britton, M.~C., 
\pedant, E.~L., \& Kulkarni, S.~R.\ 2001, \apj, 560, 390
\bibitem[Leggett, 1992]{l92} Leggett, S.K. 1992, \apjs, 82, 351 
\bibitem[Leggett, Harris, \& Dahn(1994)]{1994AJ....108..944L} Leggett, S.~K., Harris, H.~C., 
\& Dahn, C.~C.\ 1994, \aj, 108, 944
\bibitem[Leggett et al, 1996] {l96} Leggett, Allard, F., Berriman, G., Dahn, C.C., Hauschildt, P.H. 1996, \apjs, 104, 117 
\bibitem[Liebert et al]{l99} Liebert, J., Kirkpatrick, J.D., Reid, I.N., Fisher, M.D. 1999, ApJ, 519, 345
\bibitem[Luhman et al.(2000)]{2000ApJ...540.1016L} Luhman, K.~L., Rieke, G.~H., Young, E.~T., Cotera, A.~S., Chen, H., Rieke, M.~J., Schneider, G., \& Thompson, R.~I.\ 2000, \apj, 540, 1016
\bibitem[Luhman(2000)]{2000ApJ...544.1044L} Luhman, K.~L.\ 2000, \apj, 544, 1044
\bibitem[Lutz \& Upgren, 1980] {lu80} Lutz, T.E., Upgren A.R. 1980, \aj, 85, 573
\bibitem[Luyten, 1980] {luy80}  Luyten, W.J. 1980, Catalogue of stars with proper motions exceeding 
0"2 annually (NLTT), Univ. of Minnesota Publ., Minneapolis, Minnesota
\bibitem[Madau, Pozzetti, \& Dickinson(1998)]{1998ApJ...498..106M} Madau, P., Pozzetti, L., 
\& Dickinson, M.\ 1998, \apj, 498, 106
\bibitem[Madsen et al, 2002] {mdl} Madsen, S., Dravins, D., Lindegren, L. 2002, \aap, 381, 446
\bibitem[Marcy \& Benitz, 1989] {mb89} Marcy, G.W., Benitz, K.J. 1989, ApJ, 344, 441
\bibitem[Martin et al] {mal}  \pedant, E.L., Delfosse, X., Basri, G., Goldman, N.,
Forveille, T., Zapatero Osorio, M.R.  1999, \aj, 118, 2466
\bibitem[Martin(1999)]{1999MNRAS.302...59M} \pedant, E.~L.\ 1999, \mnras, 302, 59
\bibitem[Martin]{ma} \pedant, E.L., Ardila, D.R. 2001, \aj, 121, 2758

\bibitem[Mohanty \& Basri, 2002] {mb2002} Mohanty, S., Basri, G. 2002, Cool Stars 12 Workshop, 
ed. A. Brown et al.
\bibitem[Monet et al.(1992)]{1992AJ....103..638M} Monet, D.~G., Dahn, C.~C., Vrba, F.~J., 
Harris, H.~C., Pier, J.~R., Luginbuhl, C.~B., \& Ables, H.~D.\ 1992, \aj, 103, 638
\bibitem[Mould, 1976] {mou} Mould, J.R. 1976, \aap, 48, 443
\bibitem[Neugebauer et al, 1965] {n65} Neugebauer, G., Martz, D.E., Leighton, R.B. 1965, \apj, 142, 399
\bibitem[Neugebauer \& Leighton, 1969] {ng69} Neugebauer, G., Leighton, R.B. 1969, 
{\sl Two Micron Sky Survey - a preliminary catalogue}, NASA SP-3047 (Washington DC:
Government printing office)
\bibitem[Persson et al, 1998] {per98} Persson, S.E., Murphy, D.C., Krzeminski, W., 
Roth, M., Rieke, M.J. 1998, \aj, 116, 2475
\bibitem [Rebolo et al, 1992] {r92} Rebolo, R., \pedant, E.L., Magazzu, A.
1992, \apj, 389, L83
\bibitem[Reid \& Gilmore, 1981] {rg81} Reid, I.N., Gilmore, G.F. 1981, \mnras, 196, 15P
\bibitem[Reid, Tinney, \& Mould(1994)]{1994AJ....108.1456R} Reid, N., Tinney, C.~G., \& Mould, 
J.\ 1994, \aj, 108, 1456
\bibitem[Reid et al, 1995] {rg95a} Reid, I.N., Hawley, S.L., Mateo, M. 1995a, \mnras, 272, 828[RHM] 
\bibitem[Reid et al, 1995] {rg95b} Reid, I.N., Hawley, S.L., Gizis, J.E. 1995b, \aj, 110, 1838 [PMSU1]
\bibitem[Reid et al, 1999a] {r99}  Reid, I.N., Kirkpatrick, J.D., Liebert, J., Burrows, A., Gizis, J.E.,
{\sl et al.} 1999a, \apj, 521, 613
\bibitem[Reid et al, 1999] {r99b} Reid, I.N.,  Kirkpatrick, J.D., Gizis, J.E., Liebert, J. 1999b, \apjl 527, L105
\bibitem[Reid et al, 2000] {re20} Reid, I.N., Kirkpatrick, J.D., Gizis, J.E., Dahn, C.C.,
Monet, D.G., Williams, R.J., Liebert, J., Burgasser, A.J. 2000, AJ 119, 369
\bibitem[Reid \& Mahoney] {rm20} Reid, I.N., Mahoney, S. 2000, MNRAS, 316, 427 (RM2000)
\bibitem[Reid \& Hawley, 2000]{rh2000} Reid, I.N., Hawley, S.L., 2000, {\sl New Light on Dark Stars},  
Springer-Praxis (London, Berlin, Heidelberg)
\bibitem[Reid {\sl et al.}, 2001] {r201}  Reid, I.N., Gizis, J.E., Kirkparick, J.D., Koerner, D.W. 
2001, \aj, 121, 489
\bibitem [Reid \& Cruz, 2002] {rc20} Reid, I.N., Cruz, K.L. 2002, \aj, in press
bibitem[Schweitzer et al.(2001)]{2001ApJ...555..368S} Schweitzer, A., Gizis, J.~E., Hauschildt, P.~H., 
Allard, F., \& Reid, I.~N.\ 2001, \apj, 555, 368
\bibitem[Schweitzer et al.(2002)]{2002ApJ...566..435S} Schweitzer, A., Gizis, J.~E., 
Hauschildt, P.~H., Allard, F., Howard, E.~M., \& Kirkpatrick, J.~D.\ 2002, \apj, 566, 435
\bibitem[Skrutskie et al, 1997] {sk97} Skrutskie, M.F. {\sl et al.} 1997, in {\sl The Impact of Large-Scale
Near-IR Sky Survey}, ed. F. Garzon et al (Kluwer:  Dordrecht), p. 187
\bibitem[Soderblom, Duncan, \& Johnson(1991)]{1991ApJ...375..722S} Soderblom, D.~R., 
Duncan, D.~K., \& Johnson, D.~R.~H.\ 1991, \apj, 375, 722
\bibitem[Soderblom \& Mayor(1993)]{1993AJ....105..226S} Soderblom, D.~R.~\& 
Mayor, M.\ 1993, \aj, 105, 226. 
\bibitem[Soderblom et al, 1998] {s98} Soderblom, D.R., King, J.R., Henry, T.J. 1998, \aj, 116, 396
\bibitem[Stauffer et al, 1998]{stau} Stauffer, J., Schultz, G., Kirkpatrick, J.D. 1998, \apj, 499, L199
\bibitem[Stauffer et al, 1999]{stau99} Stauffer, J.~R.~et al.\ 1999, \apj, 527, 219
\bibitem[Strauss et al, 1999] {s99} Strauss, M. {\sl et al.} 1999, \apj, 522, L61
\bibitem[Tinney et al, 1993] {t93} Tinney, C.G., Mould, J.R., Reid, I.N. 1993, \aj, 105, 1045
\bibitem[Tinney, 1996] {t96} Tinney, C.G. 1996, \mnras, 281, 644
\bibitem [Tinney, 1998] {t98} Tinney, C.G. 1998, \mnras, 296, L42
\bibitem [Tinney \& Reid, 1998] {tr99} Tinney, C.G., Reid, I.N. 1999, MNRAS 301, 1031
\bibitem[Tonry \& Davis]{td79} Tonry, J., Davis, M. 1979, AJ, 84, 1511
\bibitem[Ulrich et al, 1966] {u66} Ulrich, B.T., Neugebauer, G., McCammon, D., Leighton, R.B.,
Hughes, E.E., Becklin, E. 1966, \apj, 146, 288
\bibitem[Ushomirsky {\sl et al.}, 1998] {ush} Ushomirsky, G., Matzner, C.D., Brown, E.F., Bildsten, L., 
Hilliard, V.G., Schroeder, P.C. 1998, \apj, 497, 253
\bibitem[Vogt et al] {vog}  Vogt, S.S., Allen, S.L., Bigelow, B.C.  {\sl et al.} 1994, S.P.I.E., 2198, 362
\end{thebibliography}
\end{document}